\documentclass[a4paper,12pt,epsfig]{article}

\usepackage{times}
\usepackage{amsmath}
\usepackage[T1]{fontenc}
\usepackage[latin1]{inputenc}
\usepackage{graphicx}
\usepackage{wrapfig}
\usepackage{bm}
\usepackage{here}
\usepackage{amssymb}
\usepackage{amsmath,times}
\usepackage{caption}
\usepackage{mdwlist}
\usepackage{lipsum}
\usepackage[shortlabels]{enumitem}
\usepackage{comment}
\usepackage{mathtools}
\usepackage{movie15}
\usepackage{color}
\usepackage{braket}
\usepackage[dvipsnames]{xcolor}
\usepackage{subcaption}

\usepackage[bookmarks,  
pdftitle={HEE},
pdfauthor={Emilio Rubi­n de Celis, FCEyN, UBA -IFIBA, CONICET},
pdfsubject={GB_bordes},
pdfpagelayout=OneColumn,pdfstartview=FitH]{hyperref}

\makeatletter
\newcommand{\newc}{\newcommand}
\def\bea#1\ena{\begin{align}#1\end{align}}
\renewcommand{\v}[1]{\mathbf{#1}}\renewcommand{\bm}[1]{\mbox{\boldmath{$#1$}}}
\newc{\partiald}[2]{\frac{\partial{#1}}{\partial{#2}}}
\renewcommand{\bm}[1]{\mbox{\boldmath{$#1$}}}
\newc{\D}{\displaystyle}
\newc{\noi}{\noindent}
\newc{\nn}{\nonumber}
\newc{\sen}{\sin}
\newc{\ra}{\rightarrow}
\newc{\rt}{\right}\newc{\lt}{\left}
\newc{\sqp}{\phi}
\newc{\cp}{\varphi}
\newc{\poi}{\phantom . \noi}
\newcommand{\cpdf}[1]{}
\newc{\va}{\v a}\newc{\vR}{\v R}\newc{\vr}{\v r}\newc{\vj}{\v j}\newc{\vJ}{\v J}\newc{\vB}{\v B}\newc{\vE}{\v E}
\newc{\vF}{\v F}\newc{\vk}{\v k}\newc{\vM}{\v M}\newc{\vp}{{\v p}}\newc{\vS}{\v S}\newc{\vD}{\v D}\newc{\vA}{\v A}\newc{\vH}{\v H}
\newc{\vn}{\v n}\newc{\vm}{\v m}\newc{\vv}{{\v v}}\newc{\vu}{{\v u}}\newc{\vL}{{\v L}}
\newc{\cL}{{\cal L}}
\newc{\cD}{{\cal D}}
\newc{\cP}{{\cal P}}
\newc{\mX}{{\mathcal X}}\newc{\mY}{{\mathcal Y}}\newc{\mZ}{{\mathcal Z}}
\newc{\argsinh}{{\rm{argsinh}}\,}
\newc{\vkappa}{\bm \kappa}
\newc{\vrho}{\bm \rho}
\newc{\vO}{{\bm \Omega}}
\newc{\tbsq}{{\tiny $\blacksquare\;\,$}}
\newc{\sbsq}{\small {$\blacksquare\;\,$}}
\newc{\bsq}{{$\blacksquare\;\,$}}

\providecommand{\LyX}{L\kern-.1667em\lower.25em\hbox{Y}\kern-.125emX\@}

\def\lp{\left(}
\def\rp{\right)}
\def\lbr{\left[}
\def\rbr{\right]}
\def\rk{\right\}}
\def\lk{\left\{}

\def\be{\begin{equation}}
\def\ee{\end{equation}}
\def\ba{\begin{array}}
\def\ea{\end{array}}
\def\bea{\begin{eqnarray}}
\def\eea{\end{eqnarray}}

\def\p{\partial}

\def\T{\Theta}
\def\lb{\label}
\def\d{\delta}
\def\D{\Delta}
\def\r{\rho}
\def\m{\mu}
\def\n{\nu}
\def\a{\alpha}

\def\g{\gamma}
\def\G{\Gamma}
\def\s{\sigma}

\def\l{\lambda}
\def\L{\Lambda}

\def\s{\sigma}
\def\S{\Sigma}
\def\e{\epsilon}
\def\v{\varphi}

\def\ra{\rightarrow}

\def\vr{{\bf r}}

\def\noi{\noindent}

\begin{document}

\title{Extremal Surfaces and Thin-shell Wormholes}

\author{Mariano Chernicoff$^{1,2}$, Gaston Giribet$^3$, Emilio Rub\'{\i}n de Celis$^3$}

\date{}
\maketitle

\begin{center}

$^1${ Departamento de F\'{\i}sica, Facultad de Ciencias, Universidad Nacional Aut\'{o}noma de M\'{e}xico}\\
{{\it A.P. 70-542, CDMX 04510, M\'{e}xico.}}

\smallskip

$^2${Departament de F\'isica Qu\`antica i Astrof\'isica and Institut de Ci\`encies del Cosmos (ICC), \\ Universitat de Barcelona, {\it Mart\' i Franqu\`es 1, ES-08028, Barcelona, Spain.}}

\smallskip

$^3${ Physics Department, University of Buenos Aires and IFIBA-CONICET}\\
{{\it Ciudad Universitaria, pabell\'on 1 (1428) Buenos Aires, Argentina.}}

\smallskip

\smallskip

\smallskip

\end{center}


\begin{abstract}
We study extremal surfaces in a traversable wormhole geometry that connects two locally AdS$_5$ asymptotic regions. In the context of the AdS/CFT correspondence, we use these to compute the holographic entanglement entropy for different configurations: First, we consider an extremal surface anchored at the boundary on a spatial $2$-sphere of radius $R$. The other scenario is a slab configuration which extends in two of the boundary spacelike directions while having a finite size $L$ in the third one. We show that in both cases the divergent and the finite pieces of the holographic entanglement entropy give results consistent with the holographic picture and this is used to explore the phase transitions that the dual theory undergoes.
The geometries we consider here are stable thin-shell wormholes with flat codimension-one hypersurfaces at fixed radial coordinate. They appear as electrovacuum solutions of higher-curvature gravity theories coupled to Abelian gauge fields. The presence of the thin-shells produces a refraction of the extremal surfaces in the bulk, leading to the presence of cusps in the phase space diagram. Further, the traversable wormhole captures a phase transition for the subsystems made up of a union of disconnected regions in different boundaries.
We discuss these and other features of the phase diagram.
\end{abstract}

\maketitle

\vspace{-1cm}

\newpage

\section{Introduction}
In this paper we investigate extremal surfaces in the presence of a traversable wormhole spacetime in the context of the holographic description of entanglement entropy. The bulk geometry we consider is a Lorentzian 5-dimensional wormhole that connects two asymptotic regions which are locally Anti-de Sitter (AdS). 
The wormhole is a solution of a higher-curvature gravity theory coupled to an Abelian gauge field. The gravity sector is given by the Einstein-Gauss-Bonnet (Lovelock) quadratic action, which makes the problem tractable analytically and enables us to construct the wormholes explicitly 
without extra exotic matter. The wormhole is dual to a system of a pair of interacting boundary theories given on two copies of Minkowski ${\mathbb R}^{1,3}$ spacetimes, where we investigate different entanglement entropy configurations.
\\

In the context of quantum field theory (QFT), entanglement entropy is understood as a UV-regulated quantity that measures the degree of entanglement of a quantum system, giving an estimation of a type of correlation between parts of the system that are associated to complementary spatial domains of the whole space.
Entanglement entropy thus provides important insight about the structure of the quantum state under consideration and, among other things, it allows to distinguish topological phases and characterize critical points. 
Loosely speaking, given a state of the system in a $d$-dimensional space-like Cauchy slice of a $(d+1)$-dimensional QFT, the entanglement entropy $S_A$ of the subsystem inside region $A$ on this slice is defined as the von Neumann entropy of the reduced density matrix obtained by tracing out the degrees of freedom outside region $A$, i.e. tracing over its complementary subsystem \cite{Srednicki}. For a QFT in even spacetime dimensions, the general structure of the ultra-violet (UV) behavior of the entanglement entropy is given by
\begin{equation}\label{EE}
S_{\text{A}}
=
p_1 \lp \frac{R}{\epsilon} \rp^{d-1}+ \ldots + \tilde{a} \log \lp{\frac{R}{\epsilon}} \rp+ \mathcal{O}(\e^0) \ ,
\end{equation}
where $R$ is a characteristic length scale of region $A$,
and $\epsilon$ is the UV regulator associated to some lattice spacing, such that $\e \to 0$ in the continuum limit. The leading divergent term in (\ref{EE}) is known as the {\it area law term} and its coefficient $p_1$ depends on the regularization procedure. The coefficient $\tilde{a}$ in the logarithmic term, which appears only in even spacetime dimensions, is cutoff independent and captures information about the conformal anomalies in the theory, and it plays an important role in the entanglement-based results on renormalization group flow \cite{Casini1, Myers,Liu, Casini2}.

In the context of the holographic correspondence, Ryu and Takayanagi presented a very succinct and elegant prescription to compute entanglement entropy in a great variety of strongly coupled conformal field theories (CFT) \cite{Takayanagi}. According to this prescription, the holographic entanglement entropy (HEE) for a region $A$ at the boundary is given, to leading order in the $G_{d+2}$ expansion, by the formula
\begin{equation}\label{RTHEE}
S_{\text{\tiny HEE}}^{A}
=\frac{\text{Area}(\S)}{4 G_{d+2}}\ ,
\end{equation}
where $\S$ is the $d$-dimensional minimal surface in the bulk that ends on the $(d-1)$-dimensional manifold $\partial{\S} =\partial A$ at the boundary, and $G_{d+2}$ is the Newton constant in $(d+2)$-dimensions. As stated, this result is valid when the bulk geometry is static and time independent and described by Einstein gravity in asymptotically AdS$_{d+2}$ spacetimes. Equation (\ref{RTHEE}) have passed a great number of consistency tests and as expected, when compared to (\ref{EE}) they fully agreed. A derivation of (\ref{RTHEE}) was given in \cite{Lewkowycz:2013nqa}; see also \cite{Faulkner:2013ana} and references therein and thereof. In the last few years we have also learnt how to compute the holographic entanglement entropy in a wide variety of scenarios which include for example, time-dependent geometries \cite{Hubeny}, or higher-curvature gravity, among many other setups. When higher-curvature terms are present in the gravity action, the holographic formula (\ref{RTHEE}) receives corrections; the area is replaced by the appropriate functional for the HEE, which can be systematically derived from the gravity theory, cf. \cite{Camps, Dong}. Here, being interested in the wormhole solution, we will consider the HEE generalized to 5-dimensional Einstein-Hilbert action supplemented with the quadratic Gauss-Bonnet term.
\\

The solution we consider is the one constructed in \cite{emilio}, which consists of a geometry made out of four different patches connected by three thin-shell junctions. One of such thin-shells corresponds to the wormhole throat, while the other two, as seen from each asymptotic region, can be thought of as bubbles, or the mouths of the wormhole, inside which the throat is located. The higher-curvature contributions to the boundary action makes it possible that the junctions do not contain induced matter; that is, the induced stress tensor on the thin shells, both the bubbles and the throat, vanishes. In other words, the quadratic curvature terms suffice to fulfil the generalized Israel conditions without exotic matter. In addition, the flux of a $U(1)$ gauge field suffices to support the wormhole throat and, in certain range, stabilize it. 

As studied in \cite{emilio}, the wormhole presents instabilities in the corresponding interval of the parameter space allowed by causality constraints \cite{Brigante, Brigante2}. This demands to consider a special point of the coupling, which is the so-called Chern-Simons point \cite{Zanelli}, the only point where the wormhole is stable under scalar perturbations and the arguments concluding causality violation do not apply. There, the theory exhibits special properties; among them, the absence of linear excitations around the AdS vacuum, uniqueness of such vacuum, and gauge symmetry enhancement. The wormhole geometry at the Chern-Simons point has been investigated with string probes in \cite{mariano}, where it was shown that, when going through the junctions, the string configurations produce a refraction which, from the boundary theory perspective, represents a sharp phase transition. A natural question arises as to whether a similar phase transition phenomenon occurs when probing the wormhole geometry with codimension-2 surfaces as the one considered in the holographic realization of entanglement entropy. We will show that, as it happens with the string probes, the holographic entangling surfaces also experience a refraction when going through the junctions, and from the point of view of the dual field theory this can be interpreted as a phase transition. Moreover, depending on the boundary conditions imposed on both asymptotic regions, which correspond to different entangling surfaces in each of the two interacting copies of the dual CFT, there appear extremal surfaces that extend from one asymptotic region to the other, i.e. such extremal surfaces traverse the throat of the wormhole.
A list of works considering similar setups, including similar configurations, include \cite{Myers1,deBoer,Bhattacharyya,BA2014,Mollabashi:2014qfa,He:2016fiz}; see also references therein and thereof.
\\

The paper is organized as follows: 
In Section \ref{grav_th}, we introduce the higher-curvature gravity theory that we will consider and briefly review the Chern-Simons point of its parameter space to present the wormhole solution we are going to work with. In Section \ref{HEE_GB}, we discuss the correct definition of the functional and the relevant variational problem for the holographic description of entanglement entropy in the higher-curvature theory, and deduce the matching conditions necessary to join extremal surfaces across thin-shells of the bulk spacetime. In Section \ref{HEE_examples}, we consider different entangling surfaces, corresponding to different 3-volumes in the boundary: We start with the case of a surface which is anchored at the boundary on a spatial $2$-sphere of radius $R$. Then, we also consider a slab configuration, which extends in two of the boundary space-like directions while having a finite size $L$ in the third one. For both types of configurations, we determine the extremal surfaces by carefully solving the matching conditions. Then, we compute the different contributions to the entanglement entropy by using the generalized Ryu-Takayanagi formula. 
Both the divergent and the finite pieces of the entanglement entropy will be shown to give results consistent with the holographic picture, and this fact will be used to explore the phase transitions that the dual theory undergoes. The refraction phenomenon in the bulk produced by the presence of the thin-shells will translate into the presence of cusps in the phase space diagram at the boundary theory. Furthermore, the interaction through the throat captures a phase transition associated to disconnected subsystems made up with regions in both boundaries.

\section{Gravity theory and the wormhole solution}
\label{grav_th}

Higher-curvature gravity models in 5-spacetime dimensions do allow for stable wormholes that connect two asymptotically (Anti-)de Sitter regions without introducing extra exotic matter, \cite{emilio}: We consider the Einstein-Maxwell theory supplemented with the quadratic Gauss-Bonnet higher-curvature terms; namely
\begin{equation}
I_5 = - \frac{1}{16\pi G_5 }\int \, d^5x\, \sqrt{-g}\, \Big(\, R-2\Lambda +{\alpha}\, \Big( R_{\mu \nu \rho \sigma }R^{\mu \nu \rho \sigma }-4R_{\mu \nu }R^{\mu \nu }+R^2\Big)
-\,  F_{\mu \nu} F^{\mu \nu}\, \Big) +B \ ,
\label{Uno}
\end{equation}
where $B$ stands for the boundary term that renders the variational problem well defined. The Chern-Simons point corresponds to{\footnote{The Chern-Simons point is $\lambda = 1/4$ in the notation usually considered in a number of papers that analyze higher-curvature terms in the context of AdS/CFT; cf. \cite{Brigante, Brigante2}.}} $\a \to \a_{\text{\tiny CS}} = -3/(4\L)$. 
The vacuum of the theory corresponds to AdS$_5$
\begin{equation}
ds^2 = \ell^2 
\lbr
\frac{dr^2}{r^2}+r^2(-dt^2+d\vec{x}^2)
\rbr \,,
\end{equation}
where ${t}\in \mathbb{R}$, $r \geq 0 $,
and $ d\vec{{x}}^{\,2}= \delta_{ij}d{x}^id{x}^j $ (with $i,j=1,2,3$) the metric on a locally, maximally symmetric 3-space of constant curvature $k=0$. This is the AdS$_5$ solution in Poincar\'e coordinates. The (squared) radius of AdS$_5$ is given by $\ell^2= 4\a_{\text{\tiny CS}} = -3/\L $.

As mentioned before, the theory admits stable static wormhole solutions that connect two asymptotically AdS$_5$ regions without introducing extra exotic matter. 
The latter is given by two copies of the metric
\be \lb{metric}
ds^2 = \, g_{\mu \nu }\, dX^{\mu }dX^{\nu } 
=
\ell^2 \lp - N^2 \, f(r) \, d{t}^2 + f(r)^{-1} \, dr^2 + r^2 \, d\vec{{x}}^{\,2} \rp  \ .
\ee
We will refer to each of these copies as $\mathcal{M}_{\mbox{\tiny L}}$ and $\mathcal{M}_{\mbox{\tiny R}}$, the Left and Right side, respectively. The wormhole solution is given by $\mathcal{M} = \mathcal{M}_{\mbox{\tiny L}} \cup \mathcal{M}_{\mbox{\tiny R}}$ with $\mathcal{M}_{\mbox{\tiny L}}$ and $\mathcal{M}_{\mbox{\tiny R}}$ glued together at the throat which is located at the hypersurface defined at $r=r_{\mbox{\tiny th}}$. 
Each side is constructed with an interior and an exterior region, i.e. $\mathcal{M}_{\mbox{\tiny L,R}} = \mathcal{M}^i_{\mbox{\tiny L,R}} \cup \mathcal{M}^e_{\mbox{\tiny L,R}}$, joined at the bubble surface $r=r_{\mbox{\tiny b}}$. In the Chern-Simons point, the metric functions are given by
\be  \label{f_coeff}
f(r) =
   \begin{dcases}
    {f}_i(r) = r^2 + 2\sqrt{3 - 2/r^2}
     &\mbox{in $\mathcal{M}_{\mbox{\tiny L,R}}^{i} = \{ X^\mu \, / \, r_{\mbox{\tiny th}} \le r \le r_{\mbox{\tiny b}} \}$ }
     \\
   {f}_e(r) = r^2 - 2\sqrt{{\mu} - 2/r^2} \,
&\mbox{in $\mathcal{M}_{\mbox{\tiny L,R}}^{e} = \{ X^\mu \, / \, r \ge r_{\mbox{\tiny b}} \} $} 
 \, 
   \end{dcases}
\ee
with lapse factors 
\be \label{lapse}
N^2 =
   \begin{dcases}
    N_i^2 = {f}_e(r_{\mbox{\tiny b}})/{f}_i(r_{\mbox{\tiny b}}) 
     &\mbox{in $\mathcal{M}_{\mbox{\tiny L,R}}^{i} 
     $ }
     \\
  N_e^2 = 1
&\mbox{in $\mathcal{M}_{\mbox{\tiny L,R}}^{e} 
 $} 
 \, 
   \end{dcases}
\ee
where $r_{\mbox{\tiny th}}=1$, $r_{\mbox{\tiny b}} = \sqrt{\delta} \simeq 1.40$, with $\d$ given by the positive root of the polynomial $P( \d ) =  9\d^7  - 21 \d^6  + 4 \d^5 + \d^4  + 16 \left( \d^3 - \d^2 - \d - 1 \right)$, while ${\mu} =  10/\d  - 3 - 4 /\d^{4} \simeq 1.81$. The only relevant scale is $\ell$. On the other hand, the electromagnetic field is given by $F  \, = 
F_{t r} \, dt\wedge dr$, with $F^2 = 12/(r^6\ell^2)$. We note that the exterior metric is given by a supracritically charged black brane solution of the theory; however, there are no singularities in any of the patches used for the inner and outer regions of the wormhole.
For more details on this solution see \cite{emilio} and \cite{mariano}.

With the wormhole solution at hand, we are interested in studying different aspects of the geometry in the context of the AdS/CFT correspondence \cite{Maldacena}. Motivated by the seminal work of Maldacena and Susskind \cite{Susskind}, where they conjectured that certain wormholes geometries are dual to some particular entangled states in the corresponding CFT picture, we would like to explore this direction and study probes in the wormhole configuration that could help to better understand such connection computing the holographic entanglement entropy for certain regions in the boundary field theory side. 

\section{Holographic entanglement entropy in the higher-curvature theory}
\label{HEE_GB}

\subsection{Extremal functional and Gauss-Bonnet terms}

\begin{figure}[H]
\centering
\vspace{-0.8cm}
\includegraphics[width=11cm]{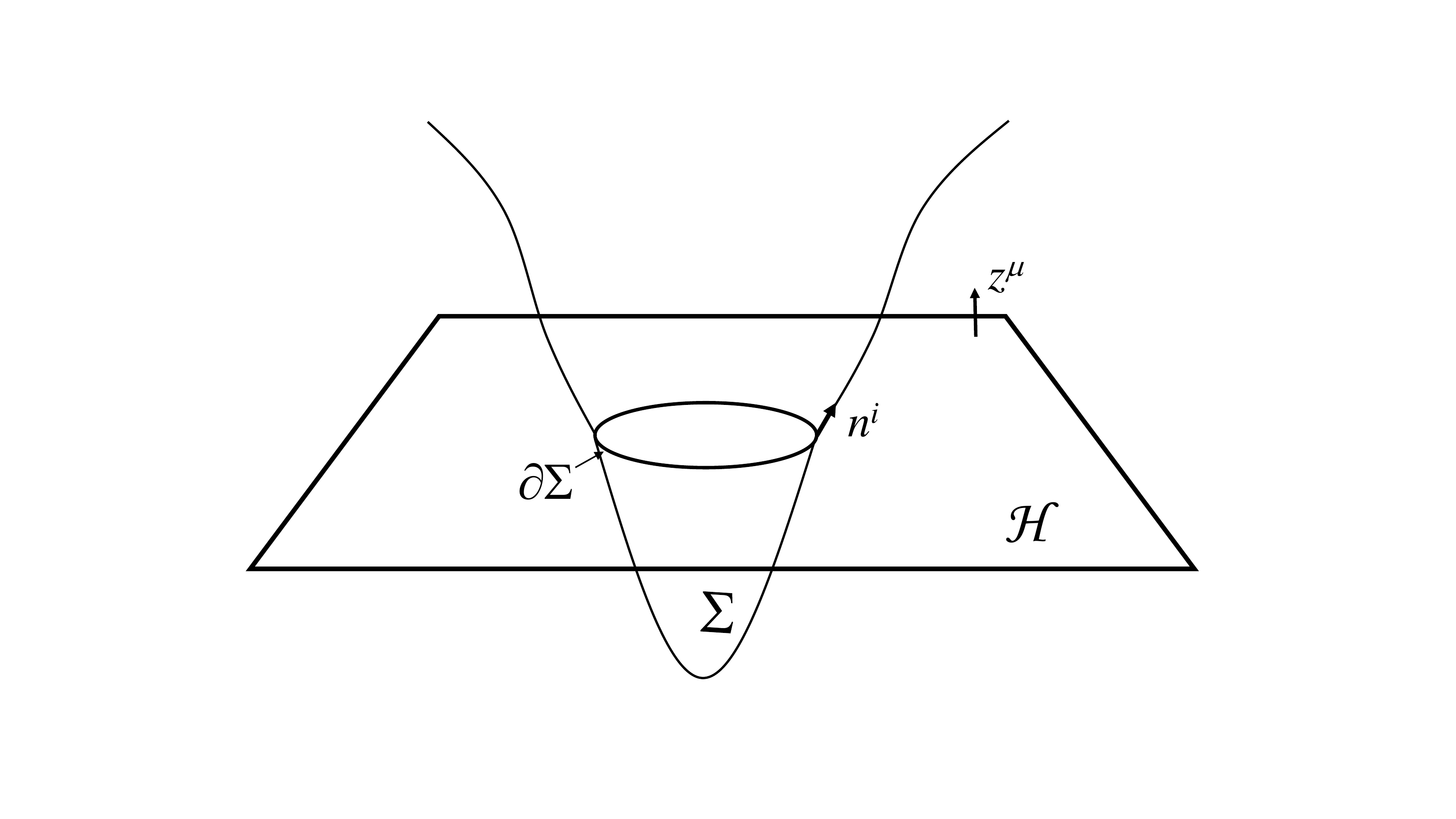}
\vspace{-1cm}
\caption{The 3-dimensional hypersurfaces $\S$ and $\mathcal{H}$ are embedded in a constant time slice of the parent bulk spacetime $\mathcal{M}$. 
Here, the hypersurface $\mathcal{H}$ with spacelike normal vector $z^\m$, represents the spacelike sheet of the parent bulk spacetime boundary $\p \mathcal{M}$. The intersection $\S \cap \mathcal{H}$ defines the 2-dimensional surface $\p \S$. The normal $n^i$ drawn above is pointing outwards with respect to the surface $\S$.
}
\label{D1}
\end{figure}

We are interested in studying 3-dimensional spacelike extremal surfaces embedded in Lorentzian solutions of higher curvature gravity. We will carry our analysis in the theory (\ref{Uno}), which is the most general theory of gravity in 5-spacetime dimensions with vanishing torsion whose fields equations are given by a rank-2 symmetric tensor, covariantly conserved and are of second order in the metric. According to \cite{Camps, Dong}, the holographic entanglement entropy from the codimension-2 hypersurface $\S$, embedded in 5-dimensional spacetime $\mathcal{M}$ with bulk coordinates $X^\mu$, $\mu=0,\dots,4$, in Einstein-Gauss-Bonnet gravity is given by
\be \label{hee}
S_{\text{\tiny HEE}}
=
\frac{1}{4G_5}
\lp
\int_{\S}
d^3 x \;
\sqrt{h} 
\lp 1 + 2 \a R \rp 
+
4\a
\int_{\p \S}
d^2 x \;
\sqrt{\g} \, 
\mathcal{K}^{(n)}
\rp \ .
\ee
%
In the above expression, $h$ and $\g$ are the determinant of the induced metric on $\S$ and its boundary $\p \S$, respectively, $R$ is the intrinsic Ricci scalar on $\S$, and $\mathcal{K}^{(n)} $ is the trace of the extrinsic curvature of $\p \S$, embedded in $\S$, with respect to the outward normal.
The induced metric over $\S$ is defined by the projector tensor
$
h_{\m \n} = g_{\m\n} + t_{\m}t_{\n} - s_{\m}s_{\n}
$, 
where $t^{\m}$ and $s^{\m}$ are the timelike and spacelike orthonormal vectors to the hypersurface, respectively. Introducing some set of coordinates $\xi^i = \{\xi^1,\xi^2,\xi^3 \}$ on $\S$ and defining the tangent vectors $e_i^\m =  \p X^\m/\p \xi^i$, such that $e_i^\m t_{\m} = 0 = e_i^\m s_{\m}$, the 3-dimensional induced metric is given as
\be
h_{ij} 
=g_{\m\n} \, \p_iX^\m \, \p_jX^\n
\ee
while the corresponding intrinsic Ricci scalar is $R = R^{ij} \, h_{ij}$. 
The projector tensor over the boundary $\p \S$ is
$\g_{\m\n}
=
h_{\m\n} - n_{\m}n_{\n}
$,
where 
$n^\mu=n^i \, \p_i X^\mu$ 
is the unit normal to $\p \S$ which is tangent to $\S$, and the trace of the extrinsic curvature is given by
$ \mathcal{K}^{(n)} = \mathcal{K}^{(n)}_{\m \n} g^{\m \n}$, with $\mathcal{K}^{(n)}_{\m \n} = n_{(\s ; \e)} \, \g^\s_{\m}\g^\e_{\n}$.
Introducing the set of coordinates
$y^a = \{y^1,y^2 \}$ on $\p \S$ and defining the tangent vectors $\tilde{e}_a^\m =  \p X^\m/\p y^a$, such that $\tilde{e}_a^\m n_\m= 0$, the 2-dimensional induced metric at the boundary $\p \S$ is written as
\bea
\g_{ab}
= g_{\m\n} \, \p_a X^\m \, \p_bX^\n 
= h_{ij} \, \p_a\xi^i \, \p_b\xi^j 
\eea
while $\mathcal{K}^{(n)} = \mathcal{K}^{(n)}_{ab} \g^{ab}$, is given by
\be \label{extrinsic_curvature}
\mathcal{K}^{(n)}_{ab}
=
n_{i;j} \, \p_a \xi^i \, \p_b \xi^j  \,,
\ee
as the extrinsic curvature of $\p \S$ with respect to the outward normal $n^i$ embedded in $\S$, as depicted in Figure \ref{D1}.

The equations of motion for the extremal surface $\S$ are obtained from $\d S_{\text{\tiny HEE}}/{\d X^\m}=0$, i.e. the variations with respect to bulk coordinates $\d X^\mu$ of the functional in (\ref{hee}) given as $S_{\text{\tiny HEE}} \equiv S[h_{ij}(X^\m(\xi^k);\g_{ab}(X^\m(y^c))]$.
Taking first the variations with respect to the metrics, $h_{ij}$ and $\g_{ab}$ correspondingly, we have
\be \label{v1}
\d S
=		
- \frac{2\a}{4\,G_5}
\lbr
\int_{\S}
d^3 x \;
\sqrt{h} \, G^{ij} \, \d h_{ij}
+
\int_{\p \S}
d^2 x \;
\sqrt{\g} \, I^{ab}_{(n)} \d \g_{ab}
\rbr
\ee
with
\be
G^{ij} = 	
R^{ij} - \frac{1}{2} \, h^{ij}\, R - \frac{1}{4\a}\, h^{ij}
\ee
where $R_{ij}$ is the intrinsic Ricci tensor, and
\be
I^{ab}_{(n)} = 	
\mathcal{K}^{ab}_{(n)} - \g^{ab} \, \mathcal{K}_{(n)} \,.
\ee
Now, considering the variations of the induced metrics; these are
$\d h_{ij}
=
\d \lp
g_{\m\n} \, \p_i X^\m \, \p_j X^\n
\rp$ 
and 
$
\d \g_{ab}
=
\d \lp
g_{\m\n} \, \p_a X^\m \, \p_b X^\n
\rp$, 
putting everything together in (\ref{v1}), and after some manipulations, the total variation of the functional is
\bea \lb{v2}
\d S
&=&
- \frac{2\a}{4G_5}
\lp
\int_{\S}
d^3 x  \;
\Xi_\mu \, \d X^\m 
- 2 
\int_{\p \S}
d^2 x \;
 \T_\mu \,
 \d \tilde{X}^\m
\rp
\eea
with
\be \label{eom}
\Xi_\mu
\equiv
\lbr 
\p_i  \lp \sqrt{h} \, G^{ij} \, \p_j X^{\n} \rp
+ 
\sqrt{h} \, G^{ij} \, \p_j X^{\s} \, \p_i X^{\r} \, \G^{\n}_{\r \s} \,  
\rbr
g_{\m\n} 
\ee
and 
\be
\T_\m
\equiv
\lbr \;\p_a \lp \sqrt{\g} \, I^{ab}_{(n)} \, \p_b X^\n \rp +\sqrt{\g} \, I^{ab}_{(n)} \, \p_a X^\s \, \p_b X^\r \, \G^\n_{\r \s} - \sqrt{\g} \, n_i G^{ij} \, \p_j X^\n\, 
\rbr 
g_{\m\n} 
\ee
and where $ \d \tilde{X}^\m$ are coordinate variations along the spatial directions of the bulk boundary $\p \mathcal{M}$.
Requiring $\d S/\d X^\m= 0$ for arbitrary variations in the bulk and with fixed boundary conditions $\p \S$, i.e. with variations $ \d \tilde{X}^\m = 0$, yields the equations of motion $\Xi_\mu
=0$ 
for the bulk hypersurface $\S$.
The projection $\Xi_\mu s^\m$ onto the spacelike normal vector to $\S$, gives the minimal surface condition
\be \lb{minsurfconf}
K + 
2\a \lp
R\, K - 2 R^{ij} \, K_{ij}
\rp
=
0\ ,
\ee
where 
\be
K_{ij}
=
s_{\m;\n} \, \p_i X^\m \, \p_j X^\n
\ee
is the extrinsic curvature of $\S$ with respect to $s^\m$ and $K = K_{ij} h^{ij}$.

\subsection{Boundary terms and matching conditions}

If the parent bulk space is constructed by gluing together different smooth geometries at some matching hypersurface, i.e. $\mathcal{M} = \mathcal{M}^- \cup \mathcal{M}^+$,  and the extremal surface $\S$ is such that it traverses this hypersurface, it is necessary to guarantee the continuity of $\S=\S^- \cup \S^+$ from one side to the other by establishing the corresponding matching conditions.
To do so we define a codimension-2 spacelike matching hypersurface $\mathcal{H}$ over which $\S^-$ and $\S^+$ are joint at their common boundary $\p \S = \p \S^{\mp}$, and write the holographic entanglement entropy functional (\ref{hee}) explicitly as
\bea \label{HEE_match}
S_{\mbox{\tiny HEE}}
=&&
\frac{1}{4G_5}
\lbr 
\int_{\S^-}
d^3 x 
\sqrt{h} 
\lp 1 + 2 \a R \rp 
+
\int_{\S^+}
d^3 x 
\sqrt{h} 
\lp 1 + 2 \a R \rp \right.
\nn \\
&& \qquad\quad + 
\left.
4\a 
\lp
\int_{\p \S^-}
d^2 x 
\sqrt{\g} \, \mathcal{K}^{(n)}
+
\int_{\p \S^+}
d^2 x 
\sqrt{\g} \, \mathcal{K}^{(n)}
\rp \rbr
+
B_{\mbox{\tiny HEE}}^{{\mbox{\tiny{fixed}}}}
\eea
with the corresponding boundary terms for $\p \S^{-}$ and $\p \S^{+}$, and with $B_{\mbox{\tiny HEE}}^{{\mbox{\tiny{fixed}}}}$ standing for other boundary terms which are typically held fixed for the variational problem under consideration. 
The matching condition is obtained by requiring $\d S_{\mbox{\tiny HEE}} = 0$, while admitting coordinate variations over the hypersurface $\mathcal{H}$ where $\p \S$ is embedded, i.e. $\d \tilde{X}^\m 
\equiv \d X^\m |_{\mathcal{H}}$, as referred in (\ref{v2}), for the corresponding boundary terms of $\p \S^{\mp}$ in (\ref{HEE_match}). 
If we use the projector tensor to $\mathcal{H}$; $H^{\m\n} \equiv \g^{\m\n} + \eta^\m\eta^\n$, 
with $\eta^\m$ as the unit normal vector to $\p \S$ which is embedded in $\mathcal{H}$, then the arbitrary variations in the matching hypersurface are $\d X^\m|_{\mathcal{H}} ={H^\m}_\n \,  \d X^\n$, and we get the continuity equations over the junction given as 
\be \label{continuity}
\lp \T_\mu^- + \T_\mu^+ \rp {H^\m}_\n  = 0 \,.
\ee 
Further, we define the normal projection $\Pi \equiv - \g^{-1/2} \, \T_\m\, \eta^\m$ to obtain the non-trivial condition for the matching. The latter is generically written as
\bea
\Pi
&=&
n^i n^j G_{ij}  \, n^{\m}\eta_\m 
+
I^{ab}_{(n)} \, \mathcal{K}^{(\eta)}_{ab} 
\lb{Pi}
\eea
where
\be \label{extrinsic_curvature_eta}
\mathcal{K}_{ab}^{(\eta)} = \eta_{\m;\n} \, \p_a X^\m \, \p_b X^\n 
\ee
is the extrinsic curvature of $\p \S$ with respect to $\eta^\m$. Finally the continuity equations (\ref{continuity}) yields the matching condition given as
\be \lb{mc}
[ \,  {\Pi} \, ] \equiv  {\Pi^+} - {\Pi^-} = 0 \,.
\ee
The relative $``-"$ sign appearing in (\ref{mc}) is adopted for the conventional definition which takes a unique orientation for the unit normal $n^\m|^{\mp}$ to $\p\S|^{\mp}$. 
Written explicitly, (\ref{mc}) reads
\bea
\lp
n^i n^j \, G_{ij}  \, n^{\m}\eta_\m
+
I^{ab}_{({n})} \, \mathcal{K}^{(\eta)}_{ab} 
\rp^-
=
\lp
n^i n^j \, G_{ij}  \, n^{\m}\eta_\m
+
I^{ab}_{({n})} \, \mathcal{K}^{(\eta)}_{ab} 
\rp^+ \,,
\eea
with ${n}^\m|^{\mp} = n^i \p_i X^\m|^{\mp}$ the purely tangential unit vector to $\S^{\mp}$ which is normal to $\p \S^{\mp}$, such that both, $n_\m^+$ and $n_\m^-$, point from $-$ to $+$.
Note , for example, that the projector tensor to $\S$ is written as $h^{\m \n}|^{\mp} = \g^{\m \n} + (n^\m\,n^\n)|^{\mp}$, from the corresponding side.

The expression (\ref{Pi}) for the normal projection $\Pi$ can be rearranged using the Gauss-Codazzi equaitons to be written in terms of 2-dimensional intrinsic and extrinsic quantities defined for the boundary junction $\p \S$, as 
\bea \lb{intr}
\Pi
&=&
- \frac{1}{2}
\lp
\mathcal{R}
+
\mathcal{K}_{(n)}^{ab} \, \mathcal{K}^{(n)}_{ab}
- 
\mathcal{K}_{(n)}^2
+
\frac{1}{2\a}
\rp
n_\m \eta^\m
+
\mathcal{K}_{(\eta)}^{ab} \, \mathcal{K}^{(n)}_{ab}
-
\mathcal{K}^{(\eta)} \, \mathcal{K}^{(n)}
\eea
with $\mathcal{R} \equiv {R}^{(2)}$ the intrinsic Ricci scalar on the surface $\p \S$, and the extrinsic curvatures given by (\ref{extrinsic_curvature}) and (\ref{extrinsic_curvature_eta}).

\begin{figure}[H]
\centering
\vspace{-.45cm}
\includegraphics[width=11cm]{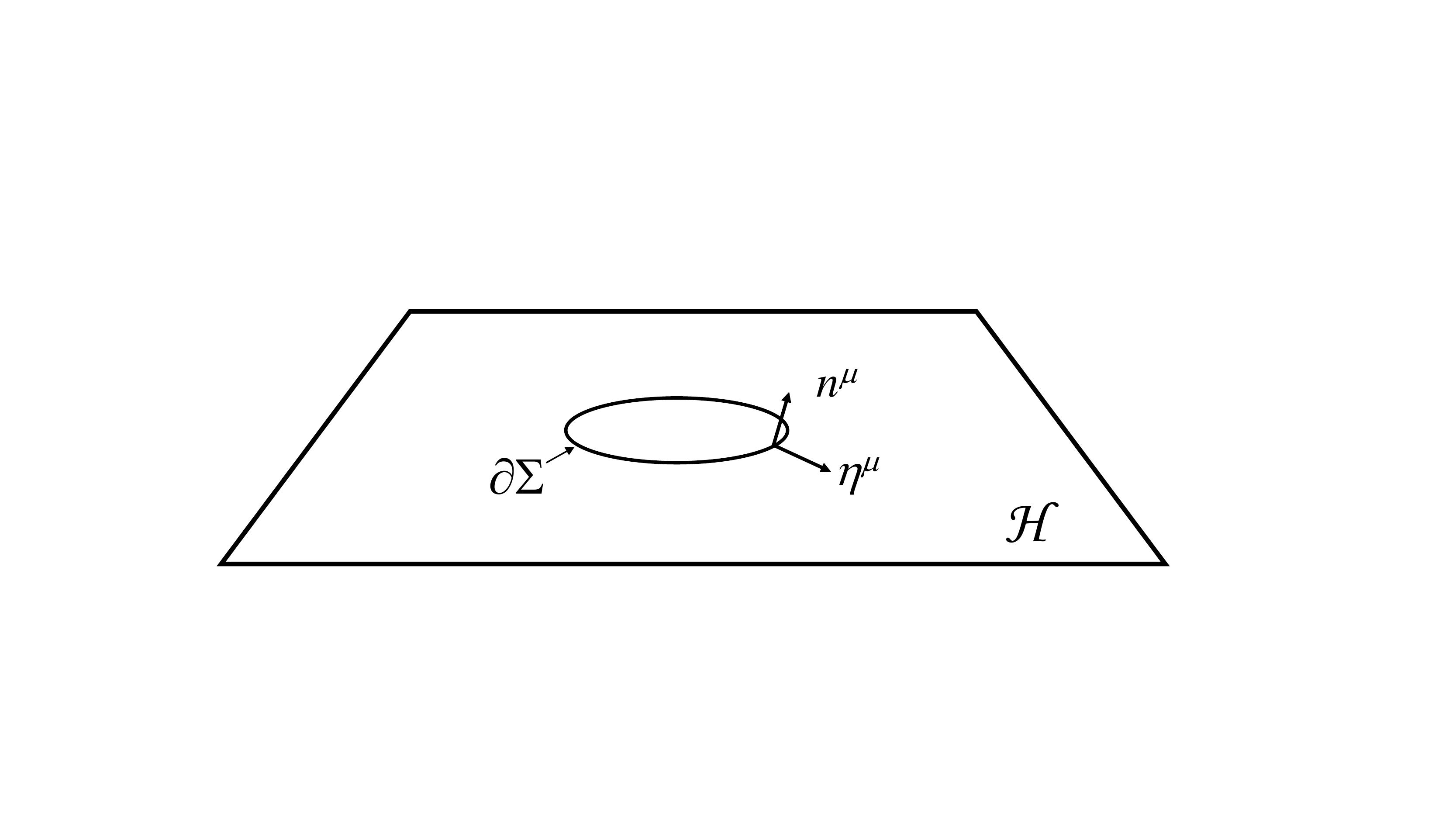}
\vspace{-1cm}
\caption{The 2-dimensional surface $\p \S$ embedded in the 3-dimensional hypersurface $\mathcal{H}$. Both are embedded in a constant time slice of the parent bulk spacetime $\mathcal{M}$.}
\label{}
\end{figure}

\section{Entanglement entropy and the Chern-Simons wormhole
}
\label{HEE_examples}

Before we begin to study our particular setup in presence of the wormhole, we compute the HEE for some known examples in pure AdS$_5$, i.e. when the wormhole is still closed. This will facilitate the comparative analysis. 
As explained in the Section \ref{grav_th}, we will consider Einstein-Gauss-Bonnet gravity at the Chern-Simons point, i.e. $\alpha = \alpha_{\text{\tiny CS}}\equiv -3/(4\Lambda)$.

\subsection{Holographic entanglement entropy in pure AdS$_5$}

In the following we consider the subsystems described as the 3-dimensional volumes of the interior of a sphere or a slab in a CFT$_4$ dual to pure AdS$_5$.

\subsubsection{A spherical volume of radius $R$}
\label{S_sphere}

\begin{figure}[H]
\centering
\includegraphics[width=10cm]{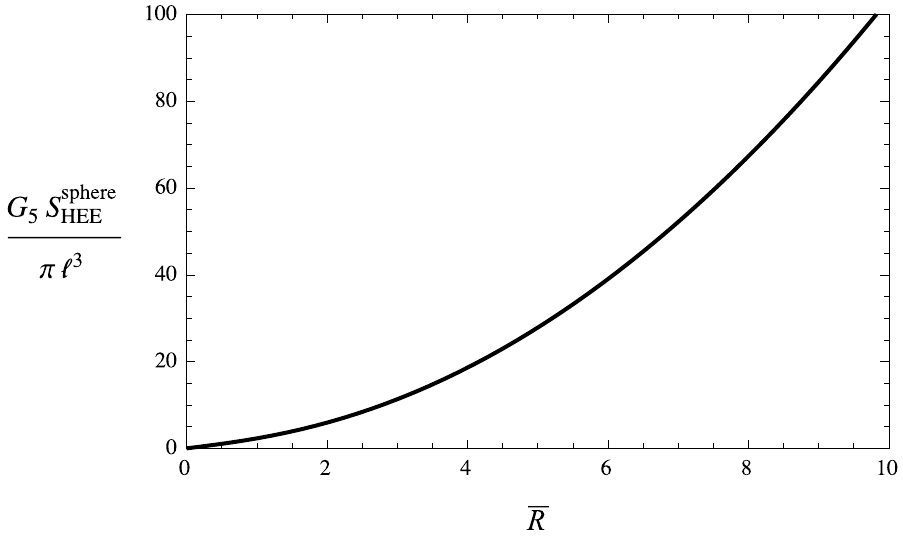}
\caption{Renormalized $S_{\text{\tiny HEE}}^{\text{sphere}}$ vs rescaled dimensionless width $\bar{R}$.}
\label{S_sphere_ren_vs_R}
\end{figure}

As a first example, let us compute the HEE when the corresponding region at the boundary is a 3-dimensional ball.
The codimension-2 surface $\S$ is embedded in pure AdS$_5$ bulk with its boundary $\partial \S$ anchored to the surface of a sphere of radius $\ell R$. 
For convenience, we use spherical coordinates, i.e. $d\vec{{x}}^{\,2}= d\rho^2+\rho^2 ( d\theta^2+ \sin^2\theta d\varphi^2)$ with $\rho \geq 0 $, $\theta\in [0,\pi]$ and $\varphi\in [0,2\pi]$. The hypersurface $\S$ can be described by the induced metric
\be
dh^2 
= \ell^2 
\lbr h_{rr}(r)\, dr^2 + r^2 \rho^2(r) (d\theta^2 + \sin^2\theta \, d\varphi^2) 
\rbr \ ,
\ee
with 
\be
h_{rr}(r) = \frac{1}{r^2} + \lp r\, {\rho'}(r) \rp^2  \,,
\ee
where $'$ stands for $\frac{d}{d r}$. 
Using (\ref{hee}), the HEE functional for the sphere
takes the following form \cite{Myers1, deBoer},
\be\label{Gaussbonnetsphere}
S_{\text{\tiny HEE}}^{\text{sphere}} 
= \frac{\ell^3}{4G_5}\, 
4\pi 
\int_{r_0}^{r_{\infty}}
dr \,
\mathcal{L}(\r(r),\r'(r),r) \,,
\ee
with $r_0$ the point where the sphere shrinks to zero size, i.e. $\rho(r_0)=0$, and $r_\infty$ is use to define the dimensionless radius of the sphere at the conformal boundary, that is $\rho(r_\infty)=R$. The coordinate $r = r_{\infty}$ plays the role of a ultraviolet cutoff, with an associated dimensionless lattice spacing at the boundary theory given by $\epsilon = 1/r_\infty$ and with the continuum limit corresponding to $r_\infty \to \infty$.
For this case, the Lagrangian density takes the following form
\bea
\mathcal{L}(\rho,\rho'(r),r)
&=&
\frac
{1 + 
2\, \rho(r)\, r^2  ( \rho(r) +  \rho'(r) \,  r)
   + 
   r^4 (r^2 \rho^2(r)+2) \rho'^2(r)}
   {r \sqrt{1 + r^4 \rho'^2(r)
   }
   } \ .
\eea
As explained in the previous section, in order to compute the HEE we first need to solve the equations of motion or, equivalently, the minimal surface condition (\ref{minsurfconf}). The corresponding solution is given by $\r(r) = (r_0^{-2} - r^{-2})^{1/2}$, 
and the evaluation of the HEE functional for the sphere yields
\bea \label{S_sphere_total}
S_{\text{\tiny HEE}}^{\text{sphere}} 
&=&
 \frac{\pi \ell^3 }{ G_5}
\lp
\frac{r_{\infty} \sqrt{r_{\infty}^2 - r_0^2}}{r_0^2} + 
\log
\frac{r_{\infty} + \sqrt{r_{\infty}^2 - r_0^2}}{r_0}
\rp
\nn \\
&\simeq &
 \frac{\pi \ell^3 }{G_5}
\lbr
\lp \frac{R}{\epsilon} \rp^2 
+ \log{\lp \frac{2 R}{\epsilon}\rp}
+ \frac{1}{2}
+ \mathcal{O}(\epsilon^{2})
\rbr \ ,
\eea
where we replaced $r_0$ in terms of the size $R$ and used $\epsilon= 1/r_\infty$, and where we considered the expansion for small lattice spacing. The UV ($r_\infty \gg 1$) divergent term in $S_{\text{\tiny HEE}}^{\text{sphere}} 
$ is
\bea \lb{S_div_spehre}
S_{\text{\tiny div}}^{\text{sphere}} 
&=&
\frac{\ell^3 }{4G_5}
\lp 
\frac{4\pi R^2}{\epsilon^2}  + 
4\pi \log \frac{1}{\epsilon}
\rp \ ,
\eea
which corresponds to the universal area law and log terms for spherical regions in a CFT$_4${\footnote{The universal logarithmic term for spherical regions from AdS$_5$ in Einstein-Gauss-Bonnet theory is given in \cite{Bhattacharyya}. One finds
\be
-4a \log{\lp \frac{\tilde{R}}{\e} \rp} \;;
\quad
a = \frac{\pi^{2} \, \mathbb{L}^3}{f_\infty^{3/2}\,\ell_P^3} (1-6 f_\infty \l)\,,
\qquad
1- f_\infty +\lambda f_\infty^2=0\,. \nn
\ee
where $\l \, \mathbb{L}^2/2 = \alpha$ and $\ell_P^3 = 8\pi G_5$. For 5-dimensional GB theory at CS point we have $\l = \l_{\text{\tiny CS}} \equiv 1/4$, 
$f_\infty=2$, $\mathbb{L}^2/f_\infty = \ell^2$,
and the universal log term coefficient is then
$a  = - \frac{\pi \ell^3}{4 G_5}$.
The universal power law terms for spherical regions of odd $d$ spatial dimensions are known as:
\bea
S_{\text{\tiny HEE}}^{\text{sphere}_d} 
+4a \log{\lp \frac{\tilde{R}}{\e} \rp}
&=&
\frac{\ell^d \, S_{d}}{4 G_{d+2} }
\lbr
p_1 \lp \frac{R}{\e} \rp^{d-1}
+
p_3 \lp \frac{R}{\e} \rp^{d-3}
+
...
+
p_{d-2} \lp \frac{R}{\e} \rp^{2}
\rbr
\eea
$$
S_{d}
=
\frac{2\,\pi^{d/2}}{\Gamma \lp d/2\rp} \,,
\quad
p_1 = p_1(d,\lambda)
\quad p_3 = p_3(d,\lambda).
$$
in our case $S_3=4\pi$ and the UV divergent term has $p_1=1$.
}}.
The finite term, independent of the UV-cutoff, is obtained as the regularized quantity 
\be \label{reg_finite}
S_{\text{\tiny finite}}^{\text{sphere}} 
= \lim_{\epsilon \to 0 
} 
\lp
S_{\text{\tiny HEE}}^{\text{sphere}} 
- S_{\text{\tiny div}}^{\text{sphere}} 
\rp
= 
\frac{\pi \ell^3 }{ G_5}
\lp 
\log{2 R}
+ \frac{1}{2}
\rp
\ee
Redefining lengths at the boundary as $\bar{R} = R/\epsilon$ to measure distances in units of lattice spacing, we can finally write
\bea \label{S_sphere_ren}
S_{\text{\tiny HEE}}^{\text{spehre}} 
&=&
 \frac{ \pi \ell^3}{G_5} \, \lp 
\bar{R} \sqrt{1+\bar{R}^2} + \log\lp{\bar{R} +  \sqrt{1+\bar{R}^2}}\rp
\rp \nn\\
&\simeq &
\frac{ \ell^3}{4G_5} \,  \lp 
4\pi\bar{R}^2  + 4\pi\log{2 \bar{R}} + \frac{4\pi}{2} +\mathcal{O}(\bar{R}^{-2})
\rp \,
\qquad (\bar{R} \gg 1) \  .
\eea
which is plotted in Figure \ref{S_sphere_ren_vs_R} showing the leading area law behaviour for large $\bar{R} = R/\epsilon$.

\subsubsection{An infinite slab of width $L$}
\label{S_SLAB}
Now, let us consider the case in which the surface is a 3-dimensional slab volume at the boundary of AdS$_5$ which has infinite size $L_\perp^2$ in the $x_2$-$x_3$ directions and width $L$ in the $x_1\equiv x$ direction. 
The boundary is placed at a large finite value of the dimensionless energy scale coordinate $r = r_{\infty}$.
%
A U-shaped codimension-2 surface $\S$ is embedded in AdS$_5$ bulk space with its boundary $\partial \S$ anchored to the surface of the slab and centered at $x=0$. The induced metric over $\S$ is
\be
dh^2 = \ell^2 \lbr h_{rr}(r)\, dr^2 + r^2 (dx_2^2+ dx_3^2) \rbr
\ee
with
\be
h_{rr}(r) = \frac{1}{r^2} + \lp r\, {x'}(r) \rp^2  \,.
\ee
The HEE functional for the slab 
is given by
\be
S_{\text{\tiny HEE}}^{\text{slab}} 
= \frac{\ell^3}{4G_5}\, 
2 \, L_\perp^2
\int_{r_0}^{r_{\infty}}
dr \,
{\mathcal{L}(x'(r),r)} \,,
\ee
where $r_0$ is the turning point of the U-shaped surface $\S$, and the one-dimensional Lagrangian density is
\be
{\mathcal{L}(x'(r),r)}
=
r^2 \sqrt{h_{rr}(r)} + \frac{1}{\sqrt{h_{rr}(r)}} \,.
\ee
Solving the equation of motion for $\mathcal{L}(x'(r),r)$ we obtain $x(r) = \sqrt{r^2 - r_0^2}/({r \,r_0})$,
and for the width $L$ of the slab at the boundary we have
\be
L 
= 
2 \int_{r_0}^{r_{\infty}}  {x'}(r)  \, dr 
=
\frac{2}{r_0} \, \sqrt{1 - r_0^2/r_\infty^2}
\,.
\ee
The evaluation of the HEE functional for the slab yields
\bea
S_{\text{\tiny HEE}}^{\text{slab}} 
&=&
\frac{\ell^3}{4 G_5} \, 2L_\perp^2
 \, r_{\infty}^2 \sqrt{1 - r_0^2/r_{\infty}^2}  \nn \\
&\simeq &
\frac{ \ell^3}{4 G_5}  \,
\lbr
2 \lp\frac{L_{\perp}}{\epsilon}\rp^2 - 4\lp\frac{L_{\perp}}{L}\rp^2
+ \mathcal{O}(\epsilon^2)
\rbr
\label{S_slab_total}
\eea
where in the second step we replaced $r_0$ in terms of the width $L$ and used $\epsilon=1/r_\infty$, to show the expansion for small lattice spacing. 
In the continuum limit, $\epsilon \to 0$, $S_{\text{\tiny HEE}}^{\text{slab}}$ diverges as 
\bea \lb{S_div_slab}
S_{\text{div}}^{\text{slab}} 
&=&
\frac{\ell^3}{4 G_5}
\frac{2L_{\perp}^2}{\epsilon^2}
\eea
corresponding to the universal area law divergence for slab-like regions in the CFT$_4$.
The finite term, independent of the UV-cutoff, is obtained as the regularized quantity 
\be \lb{S_finite_slab}
S_{\text{finite}}^{\text{slab}} 
= 
\lim_{\epsilon \to 0 
} \lp S_{\text{\tiny HEE}}^{\text{slab}} - S_{\text{div}}^{\text{slab}}  \rp
= 
- \frac{\ell^3}{4 G_5} \frac{ 4L_{\perp}^2}{L^2} \,.
\ee
Redefining lengths at the boundary as $\bar{L} = L/\epsilon$ and $\bar{L}_\perp = L_\perp/\epsilon$ to measure distances in units of lattice spacing, we can write
\bea
S_{\text{\tiny HEE}}^{\text{slab}} 
&=&
\frac{\bar{A}\, \ell^3}{4 G_5}
\frac{\bar{L} }{\sqrt{4 + {\bar{L}}^2}} 
\label{S_slab_ren} \nn \\
&\simeq&
\frac{\bar{A}\, \ell^3}{4 G_5}	 
\lbr
1 - \frac{2}{\bar{L^2}}
+ \mathcal{O}\lp \bar{L}^{-4} \rp
\rbr \qquad (\bar{L} \gg 1)
\eea
with $\bar{A} = 2\bar{L}_\perp^2$ the rescaled dimensionless area of the slab given as the sum of its two infinite faces. 

\begin{figure}[H]
\centering
\includegraphics[width=10cm]{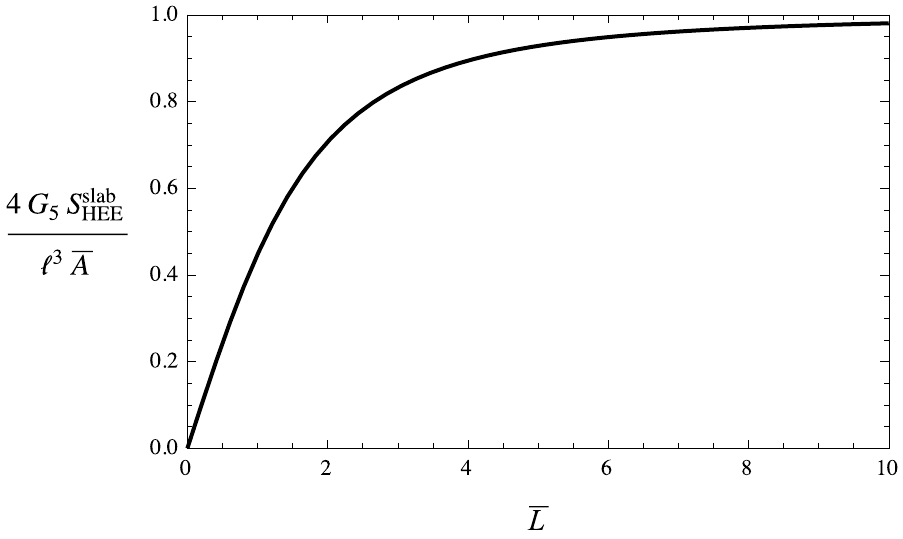}
\caption{Renormalized $S_{\text{\tiny HEE}}^{\text{slab}}$ vs rescaled dimensionless width $\bar{L}$.}
\label{S_slab_ren_vs_L}
\end{figure}

\subsection{Holographic entanglement entropy in the traversable wormhole geometry}

Now, let us open the wormhole: We will calculate the HEE hypersurfaces in the traversable wormhole geometry (\ref{metric}), which would be dual to two interacting boundary theories. We consider the volume of 3-dimensional balls and slabs in one or both boundary regions of the wormhole to probe the bulk background with the corresponding hypersurfaces $\S$. As explained before, each side of the wormhole has an inner region and an outer region separated by a bubble surface located at $r=r_{\mbox{\tiny b}}$. The Left and Right side of the wormhole join at the throat which is placed at $r=r_{\mbox{\tiny th}}<r_{\mbox{\tiny b}}$ in the holographic coordinate of each side. A hypersurface $\S$ dipping from the asymptotic boundary into the inner region gets refracted when crossing the bubble, and, if it also traverses the wormhole, then it gets refracted also when crossing the throat. These refraction effects are determined by the matching condition (\ref{mc}), which guarantee the continuity of $\S$ across a bubble, or the throat. Considering this we find solutions by constructing the minimal hypersurfaces in the wormhole geometry to compute the HEE and study the possible configurations from the perspective of the dual gauge theory.

\subsubsection{HEE in the wormhole geometry: A spherical volume of radius $R$}

The spherical region of dimensionless radius $R$ is placed at a constant time slice of one of the AdS boundaries of the wormhole geometry (\ref{metric}) located at $r=r_\infty$. We use spherical coordinates to describe the fixed-$r$ 3-space slices of constant curvature $k=0$, i.e. $d\vec{{x}}^{\,2}= d\rho^2+\rho^2 ( d\theta^2+ \sin^2 \, \theta d\varphi^2)$. In this way, the holographic hypersurface $\S$ which is anchored at the sphere of radius $R$ is embedded on the bulk geometry with an indued metric given by the line element
\be
dh^2 
= \ell^2 
\lbr 
h_{rr}(r)\, dr^2 + r^2 \rho^2(r) (d\theta^2 + \sin^2\theta \, d\varphi^2) 
\rbr 
\ee
where
\be
h_{rr}(r) = \frac{1}{f(r)} + \lp r\, {\rho'}(r) \rp^2  \,,
\ee
with $\r(r)$ as a function of $r \in [r_0, r_\infty]$, and the radius of the volume at the boundary defined as $\ell R = \ell \rho(r_\infty)$. 

\begin{figure}[H]
\centering
\includegraphics[width=11cm]{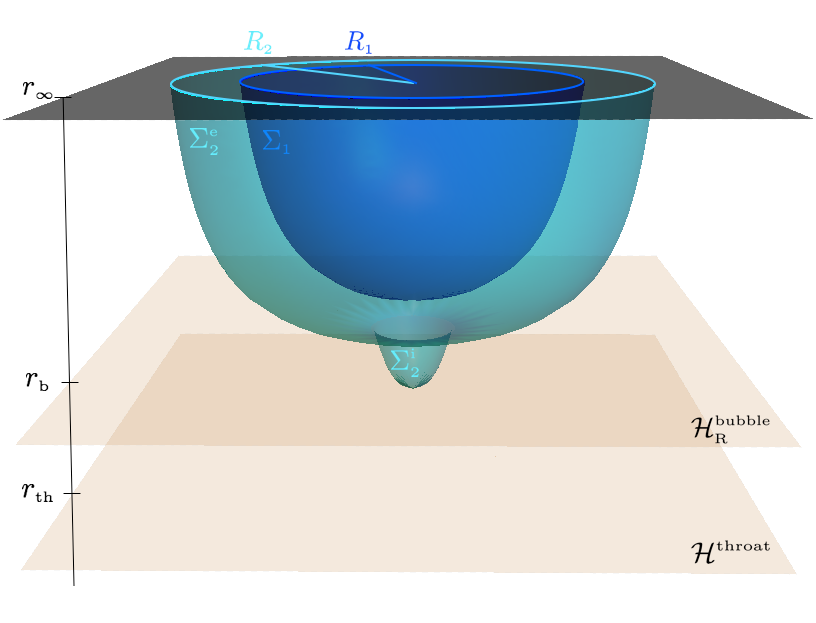}
\caption{Shape of two different spherical hypersurfaces: $\S_1$ (in blue) is in the outer region,  $\S_2  =\S_2^{\mbox{\tiny i}} \cup \S_2^{\mbox{\tiny e}}$ (in cyan) is refracted to the inner region at the Right bubble hypersurface $\mathcal{H}_{\mbox{\tiny R}}^{\mbox{\tiny bubble}}$.
}
\label{Sphere_shape}
\end{figure}

\begin{figure}[H]
\centering
\includegraphics[width=8cm]{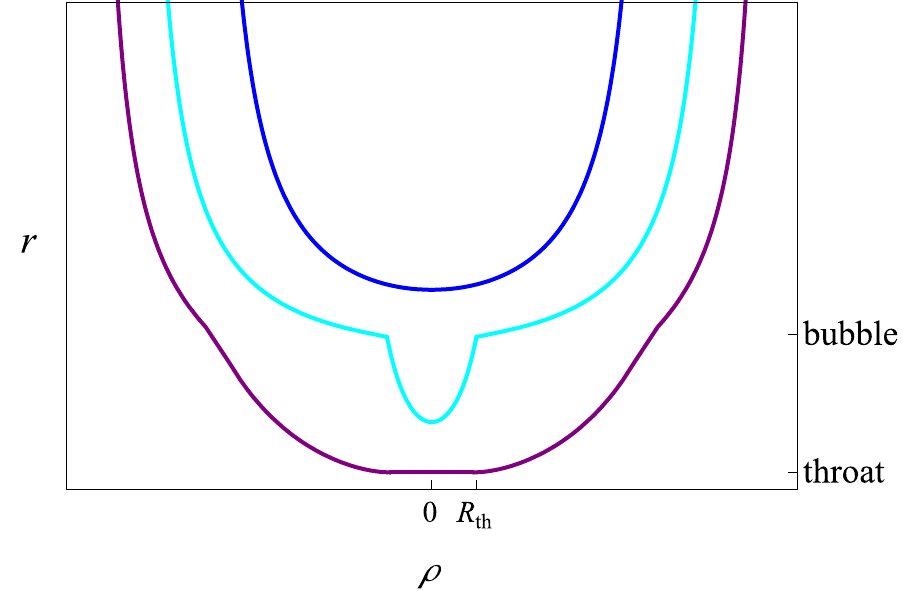}
\hspace{0.5cm}
\includegraphics[width=8cm]{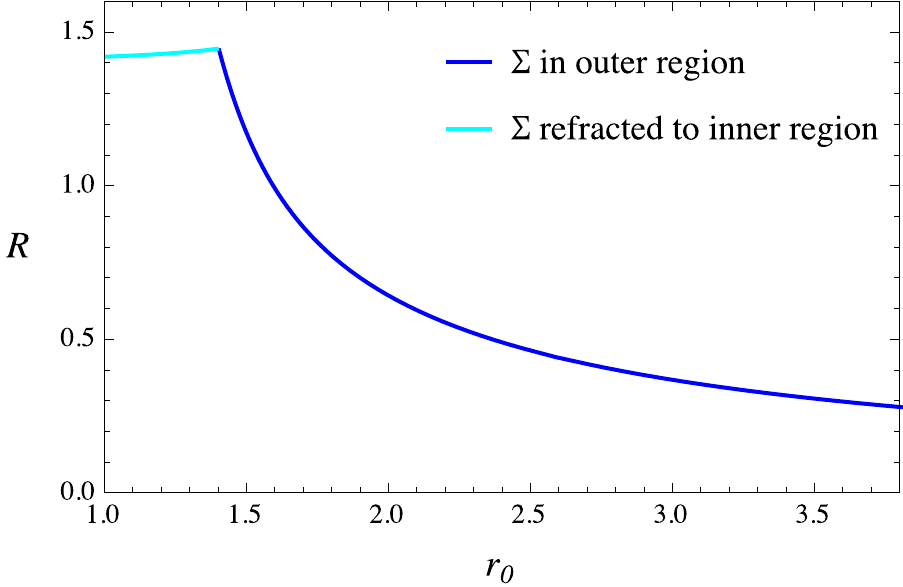}
\caption{Left: shape of different spherical hypersurfaces $\S$; in the outer region (blue), refracted to the inner region (cyan), and a stretched solution (purple). Right: radius $R$ vs. shrinking point $r_0$.}
\label{solutions_sphere}
\end{figure}

If a hypersurface, for instance $\S_1$, has its shrinking point $r_0$ in the outer region of the wormhole geometry, then it is described by the exterior parameters of the metric functions (\ref{f_coeff}) and (\ref{lapse}). If a hypersurface, say $\S_2$, shrinks in the inner region, i.e.  $r_0 < r_{\mbox{\tiny b}}$, then the interior and exterior metric coefficients are used as appropriate; we will have $\r_e(r)$ for the outer solution and $\r_i(r)$ for the inner part. The latter corresponds to the aforementioned refracted hypersurfaces $\S_2= \S_2^{\mbox{\tiny i}} \cup \S_2^{\mbox{\tiny e}}$. An example of both, $\S_1$ and $\S_2$, are shown on Figure \ref{Sphere_shape} where in the vertical direction we put the holographic coordinate $r$ of the Right side of the wormhole geometry, and in the perpendicular planes we represent $\r(r)$-$\varphi$ slices with fixed $\theta$.
To construct the refracted solutions we need to find the minimal hypersurface (\ref{minsurfconf}) in each of the bulk regions and apply the junction condition (\ref{mc}) 
to match the surfaces at the common boundary (e.g., at a bubble). Particularly, to characterize $\S$, the spacelike unit normal to the hypersurface is generically written as
\be
s_{\m}  =
\frac{\ell \, r }{\sqrt{f(r) \, h_{rr}(r)}} 
\lp
- \rho'(r) \, \d_\m^r 
+ \d_\m^\r
\rp \,,
\ee
while the time-like unit normal is given by $t_{\m}  = \ell N \sqrt{f(r)} \, \d_\m^t \,$. To establish matching conditions we define $\p \S_{\mbox{\tiny R}}$ as the intersection of $\S$ with a codimension-2 hypersurface $\mathcal{H}$ of the wormhole background at constant time and fixed $r$. The latter hypersurface is typically a bubble placed at $r=r_{\mbox{\tiny b}}$, or the throat at $r=r_{\mbox{\tiny th}}$, with normal vectors $t^\mu$ and $z^{\mu}= \ell^{-1} \sqrt{f(r)} \, \d_r^\mu$. Hence, the induced metric over $\p \S_{\mbox{\tiny R}}$ has line element
\be \label{borde_pegado}
d\g^2 = 
\ell^2 r^2 \, \rho^2(r)  
\lp 
d\theta^2 + \sin^2\theta \, d\varphi^2
\rp \,,
\ee
while  the unit normal to $\p\S_{\mbox{\tiny R}}$ which is tangent to $\mathcal{H}$ is
\be
\eta_{\m} 
= \ell \, r \, \d_\m^\r \,.
\ee
The latter is used to project over $\mathcal{H}$ to set the matching condition, alike the 2-dimensional extrinsic curvature of $\p \S_{\mbox{\tiny R}}$ with respect to $\eta^\mu$, given by
\be
\mathcal{K}^{(\eta)}_{\mu \nu }
=
\ell \, r \, \r(r) \,
\lp
\d^{\theta}_\mu \, \d^{\theta}_\nu
+
\d^{\varphi}_\mu \, \d^{ \varphi}_\nu \, \, \sin^2\theta
\rp \,,
\ee
and its trace
\be
\mathcal{K}^{(\eta)} 
=
  \frac{2}{\ell \, r \, \r(r) } \,.
\ee
The previous quantities are evaluated at the refractive surface, placed at $r=r_{\mbox{\tiny b}}\simeq 1.40$ if it is a bubble, or at $r=r_{\mbox{\tiny th}}=1$ if it is at the throat.
On the other hand, to obtain the remaining ingredients needed for the matching, we write the unit normal to $\partial \S_{\mbox{\tiny R}}$ which is tangent to $\S$ as
\be
n_{\m}  
=
\frac{\ell}{f(r) \,\sqrt{h_{rr}(r)}} 
\lp
{ \d^r}_\m 
+  r^2\, f(r)\, \rho'(r) \, {\d^\r}_\m
\rp%
\,,
\ee
pointing in the increasing direction of the bulk coordinate $r$, i.e. if $n^{\m}$ is defined on the Right (Left) side $\mathcal{M}_{{\text{\tiny R(L)}}}$ of the wormhole it points in the Left (Right) to Right (Left) direction. The 2-dimensional extrinsic curvature of $\p \S_{\mbox{\tiny R}}$ with respect to $n^\m$ is
\be
\mathcal{K}^{(n)}_{\m \n}
=
 \frac{\ell \, r  \, \r(r)}{\sqrt{h_{rr}(r)}}
\, [\r(r)  + r \r'(r)]
\lp
\d^{\theta}_\m \, \d^{\theta}_\n 
+
\d^{\varphi}_\m \, \d^{\varphi}_\n \, \sin^2\theta  
\rp
\ee
and its trace is given by
\be
\mathcal{K}^{(n)} 
=
  \frac{2 \lbr \r(r)  + r \r'(r) \rbr}{\ell \, r \, \r(r) \, \sqrt{h_{rr}(r)}} \,.
\ee
Finally, to write the matching condition $[\Pi] = \Pi^+ - \Pi^- =  0$ we need to compute the 2-dimensional intrinsic Ricci scalar of $\p \S_{\mbox{\tiny R}}$ using the induced metric (\ref{borde_pegado}), this is $\mathcal{R}=2/(r\r(r)\ell)^{2}$, and put everything together to have the general expression of the projection $\Pi$ for spherical configurations\footnote{The orientation of the normal $n^\mu$ used for this expression is given by the increasing direction of the $r$ coordinate. The reverted orientation is obtained by an overall sign difference.}
\be \label{Pi_sphere}
\Pi 
= - \frac{
2 \r(r) 
+ 
   r \,  \r'(r) \, \lp \r^2(r) (r^2 - f(r)) + 3 \rp
+ 
   r^3 f(r) \, \r'^3(r)\, (2 + r^2 \r^2(r) )
     }
     {\ell^2\, r^2 \r^2(r) f(r) \, h_{rr}^{3/2}(r)} \,.
\ee
To obtain a refracted solution which shrinks in the inner region we apply $[\Pi] = 0$ evaluated at both sides of $r=r_{\mbox{\tiny b}}$ to match $\rho_i(r)$ in the interior region with $\rho_e(r)$ in the exterior; more precisely, being $\rho_i(r_{\mbox{\tiny b}}) = \rho_e(r_{\mbox{\tiny b}})$, the matching conditions gives the relation between $\rho'_i(r_{\mbox{\tiny b}})$ and $\rho'_e(r_{\mbox{\tiny b}})$. An explicit example of a refracted solution is plotted in cyan on the left side of Figure \ref{solutions_sphere}. The minimum possible value for a shrinking point is the throat; the geometry does not admit traversing $\S$ configurations which go through the throat from one side and shrink on the other side of the wormhole. Another spherical configuration which is possible is the ball solution which lie down tangent to the throat at $r=r_{\mbox{\tiny th}}$. The latter exists because the wormhole geometry satisfies the flare-out condition at $r=r_{\mbox{\tiny th}}$, i.e. the volume of a hypersurface at the throat is minimum. 
A ball-like solution can be cut at any fixed value $R_{\mbox{\tiny th}}$, to have a ball hypersurface with $\r \in [0, R_{\mbox{\tiny th}}]$ at $r_{\mbox{\tiny th}}=1$, and matched smoothly to an inner solution with $\r_i(r) \geqslant \rho_i(r_{\mbox{\tiny th}}) = R_{\mbox{\tiny th}}$ if we set $\rho_i'(r_{\mbox{\tiny th}}) \to \infty$ to fulfill the matching. Further, the inner part is matched to an outer solution at the bubble so as to construct a stretched configuration as the union of the three parts: $\S^{\mbox{\tiny stretched}} = \S^{\mbox{\tiny ball}} \cup \S^{\mbox{\tiny i}} \cup \S^{\mbox{\tiny e}}$, an example is shown in purple in Figure \ref{solutions_sphere}. In addition to solutions with infinite $\rho_i'(r_{\mbox{\tiny th}})$, the matching condition also allows stretched solutions with finite $\rho_i'(r_{\mbox{\tiny th}})$, i.e. non-smoothly joined at $\rho_i(r_{\mbox{\tiny th}}) = R_{\mbox{\tiny th}}$, within the range $0 < R_{\mbox{\tiny th}}\lesssim 5.35$. The physical configuration among outer, refracted, or stretched solutions will be determined by the smaller value of the HEE.
On the right side of Figure \ref{solutions_sphere} we plot the dimensionless radius $R=\r_e(r_\infty)$ of a sphere at the boundary in terms of the shrinking point of the hypersurface $\S$ which is represented with the holographic dimensionless energy scale coordinate $r_0$. The blue line shows that $R$ increases as long as $r_0$ sinks deeper in the outer region; for shrinking points kept far from the bubble -with $\S$ probing only the ultraviolet region- we have $R\sim 1/r_0$ as in pure AdS, for lower $r_0$ then the radius $R$ grows faster than what is found in the pure AdS geometry. If $\S$ is in the inner region the behavior of $R$ with $r_0$ is inverted; the radius of the spherical region at the boundary decreases from $R \simeq 1.45$ to $R \simeq 1.42$ as the shrinking point dips from the refractive bubble towards the throat in the infrared sector, as shown in the cyan plot of the right side of Figure (\ref{solutions_sphere}).  On the other hand, computing for stretched configurations, it is found that the radius $R$ increases almost linearly with the radius of the ball segment stretched at the throat or, more precisely, with the value of $R_{\mbox{\tiny th}}$. The small departures from the linear relation are found for small $R_{\mbox{\tiny th}}$ in virtue of the effects of the specific matching at the throat on the computation of the integral of $\r'(r)$ along the holographic coordinate from $r_{\mbox{\tiny th}}$ to $r_\infty$. Nevertheless, for large values of $R_{\mbox{\tiny th}}$ the size of the ball at the boundary is given almost entirely by the stretching, i.e. $R \simeq R_{\mbox{\tiny th}}$. 
\\

\begin{figure}[H]
\centering
\includegraphics[width=10.5cm]{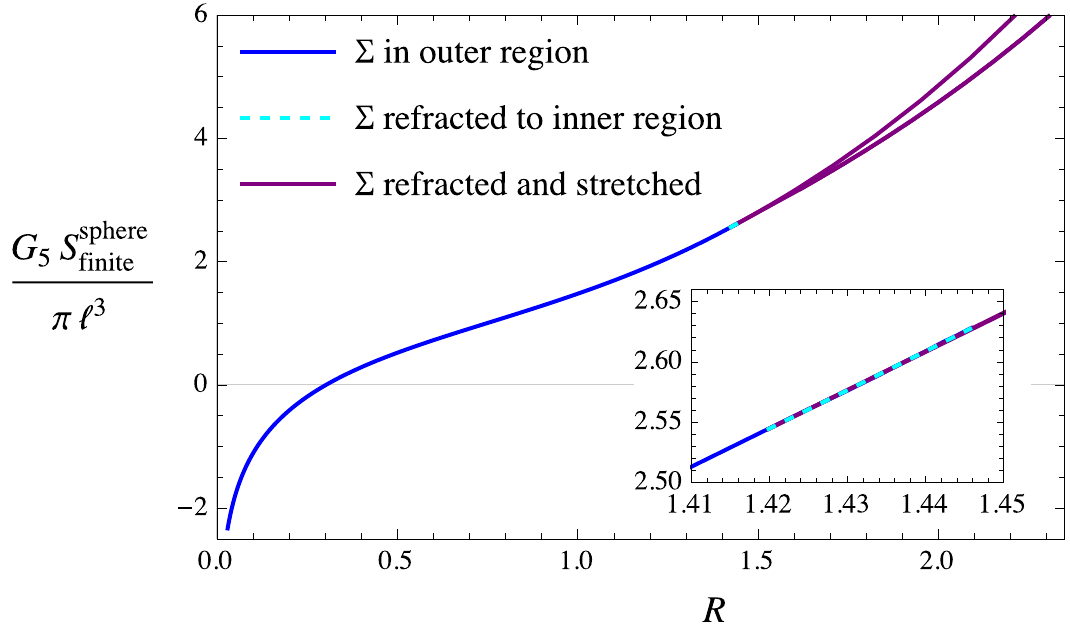}
\vspace{-0.25cm}\caption{$S_{\text{\tiny finite}}^{\text{sphere}}$ vs. radius $R$ of the subsystem given by a ball of radius $\ell R$ on one boundary.} 
\label{S_sphere_finite}
\end{figure}

The HEE functional for the sphere in the wormhole geometry takes the following form
\be \lb{S_sphere}
S_{\text{\tiny HEE}}^{\text{sphere}} 
= \frac{\ell^3}{4G_5}\, 
4\pi 
\int_{r_0}^{r_\infty}
dr  \,
\mathcal{L}(\rho,\rho'(r),r)
\ee
with the Lagrangian density
\bea
\mathcal{L}(\rho,\rho'(r),r)
&=&
\frac
{1 + (f(r)+r^2) \rho^2(r) + 
   2 r f(r) \rho(r) \rho'(r) + 
   r^2 f(r) (r^2 \rho^2(r)+2) \rho'^2(r)}
   {f(r) \sqrt{h_{rr}(r) 
   }
   } 
\eea 
evaluated in the corresponding -inner or outer- region. In the case of a stretched configuration we have 
\bea
\lb{S_sphere_str} 
S_{\text{\tiny HEE}}^{\text{sphere (stretched)}} 
&=& \frac{\pi\, \ell^3 }{G_5}\, 
\lp
\int_{r_{\mbox{\tiny th}}}^{r_\infty}
dr  \,
\mathcal{L}(\rho,\rho'(r),r)
+
r_{\mbox{\tiny th}} \int_{0}^{R_{\mbox{\tiny th}}}
2+ r_{\mbox{\tiny th}}^2 \, \rho^2
\, d\rho
\rp
 \nn \\
&=& \frac{\pi\, \ell^3 }{G_5}\, 
\int_{1}^{r_\infty}
dr  \,
\mathcal{L}(\rho,\rho'(r),r)
+
 \frac{\ell^3 }{4G_5}\, 
\lp
8 \pi\, R_{\mbox{\tiny th}} + \frac{4\pi}{3} \,R_{\mbox{\tiny th}}^3
\rp
\eea
with $R_{\mbox{\tiny th}} = \r_i(r_{\mbox{\tiny th}})$ the radius of the ball segment $\S^{\mbox{\tiny ball}}$ stretched at the throat. 
In every case, for asymptotic values of $r$, the integrand behaves as 
$\mathcal{L}(\rho,\rho'(r),r) \sim
2 \, R^2 \, r + 1/r$, producing the expected UV-divergent terms 
$S_{\text{\tiny div}}^{\text{sphere}}$,
as in (\ref{S_div_spehre}).
The finite terms, independent of the UV-cutoff, are obtained as defined in (\ref{reg_finite}), this is the regularized quantity 
\be \lb{finite_sphere_wh}
S_{\text{\tiny finite}}^{\text{sphere}}
= 
\lim_{\e \to 0} 
\lp
S_{\text{\tiny HEE}}^{\text{sphere}}- S_{\text{\tiny div}}^{\text{sphere}}
\rp
\ee
with $\e = 1/r_\infty$. The  result of the computation of (\ref{finite_sphere_wh}) is shown in Figure \ref{S_sphere_finite} where we plotted the cutoff-independent finite part of the HEE against the radius $R$ for the spherical configurations described above. The blue curve represent solutions probing the outer region of the wormhole; for $R \ll 1$ we find $S_{\text{\tiny finite}}^{\text{sphere}} \sim \log{2R} + 1/2$, while for $R \sim 1$ there is a change in the concavity with a linear behavior up to the larger outer solution for which $R\simeq 1.45$. The refracted extremal surfaces which are plotted with a cyan dashed-curve show that $S_{\text{\tiny finite}}^{\text{sphere}}$ decreases with decreasing $R$, i.e., as the shrinking point approaches the throat. 
The small range of $R$ radii covered by these refracted solutions is enlarged in the corner of the figure. 
For stretched configurations, shown in purple in Figure \ref{S_sphere_finite}, we found that the finite HEE increases with $R$ almost superimposed with the previous curves as long as the stretching is small, this is for $R_{\text{\tiny th}} < 1$. For larger radius the extensive behavior $S_{\text{\tiny finite}}^{\text{sphere}}  \sim R_{\text{\tiny th}}^3 \simeq R^3$ is manifested. 
Precisely, for a large ball of volume $V_3={4\pi}(\ell R)^3/{3}$ at the boundary, we have 
\be
S_{\text{\tiny finite}}^{\text{sphere}} \simeq  \frac{V_3}{4G_5} 
\,\qquad  (R \gg 1)\,.
\ee
If the radius $R$ is taken to infinity, $\S$ has no boundaries and becomes just the throat with infinite hypersurface $V_3^{\text{\tiny th}}$, then, correspondingly we have 
\be
S_{\text{\tiny HEE}}^{\text{Right boundary}} 
=
S_{\text{\tiny HEE}}^{\text{Left boundary}} 
=
\lim_{R \to \infty} S_{\text{\tiny HEE}}^{\text{sphere}} 
=  
\frac{V_3^{\text{\tiny th}}}{4G_5} \,
\ee
for the entanglement between the Right boundary and its complement, the Left boundary.

Finally we point out that Figure \ref{S_sphere_finite} shows two different branches of stretched configurations; the lower HEE corresponds to solutions which match $\r_i'(r_{\text{\tiny th}})$ smoothly at the throat, while the larger one corresponds to solutions with a non-smooth match of $\r_i'(r_{\text{\tiny th}})$ with the stretched segment. 
We can also note that for the small interval of radiii $R$ where the transition from outer to stretched hypersurfaces appear, the spherical regions with endpoints at one boundary have associated extremal surfaces corresponding to the three types of configurations in the bulk. Among these, the refracted solutions which shrink in the inner region are physically less favored as they have a slightly larger value of HEE in comparison with either an outer, or stretched, configuration in the same interval of the $R$ parameter. 
For regions defined in one of the boundary gauge theories 
the transition from one type of configuration to another occurs at the length scale associated to the change in concavity of the physical curve of finite HEE against $R$. 
In the following example we study the HEE for a disconnected subsystem given by two spherical regions in different boundaries to explore a transition among joint and disjoint configurations in the traversable wormhole geometry.

%

\subsubsection{HEE in the wormhole geometry: Two spherical volumes of radius $R$}

\begin{figure}[H]
\centering
\includegraphics[width=8.4cm]{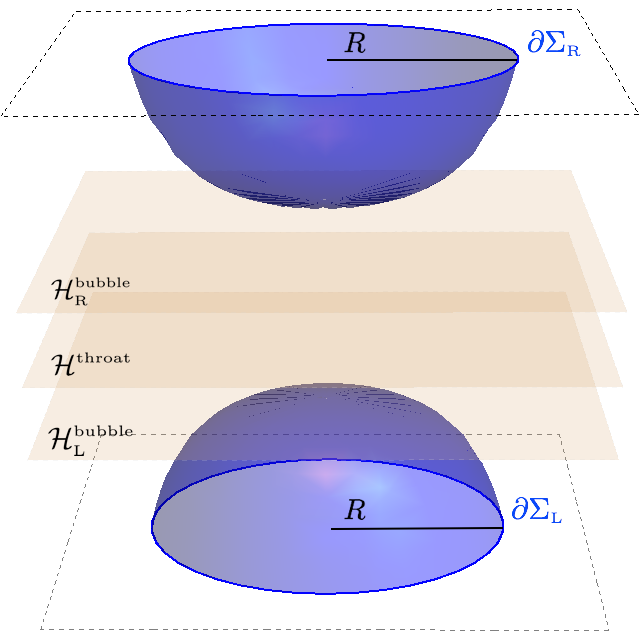}
\hspace{0.5cm}
\includegraphics[width=8.4cm]{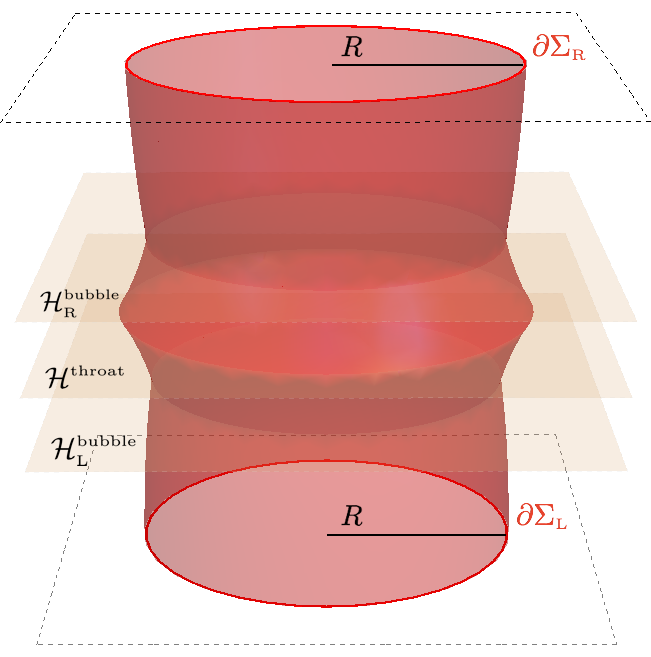}
\caption{Left, in blue: Left and Right side disjoint spherical hypersurfaces in the outer regions of the bulk wormhole geometry. Right, in red: Left-Right spherical hypersurface traversing the wormhole throat.}
\label{LRsphere}
\end{figure}

\begin{figure}[H]
\centering
\includegraphics[width=10.5cm]{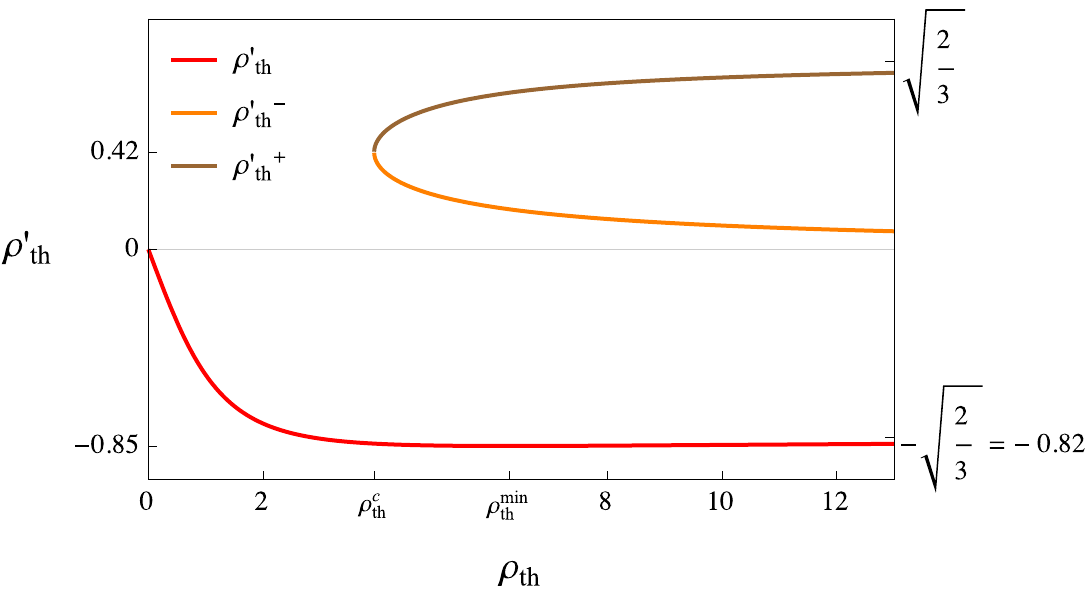}
\caption{The value of $\rho'_{\mbox{\tiny th }}$ in terms of the size $\rho_{\mbox{\tiny th }}$ of the surface $\S$ at the throat of the bulk geometry for the three possible solutions which admit configurations which traverses the wormhole throat.}
\label{solsLR}
\end{figure}

\begin{figure}[H]
\centering
\includegraphics[width=10.5cm]{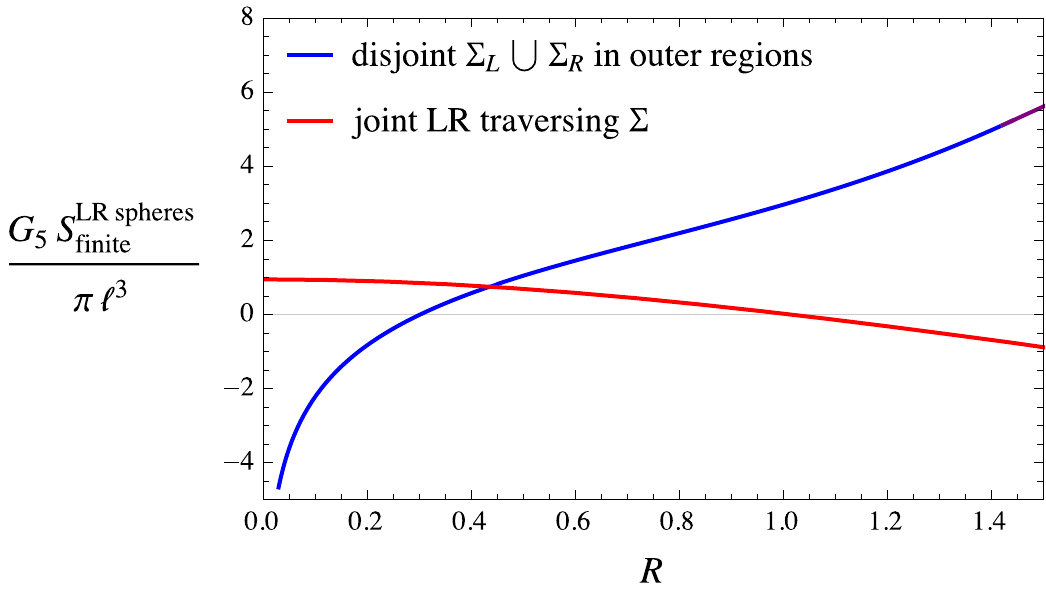}
\caption{
Finite part of the HEE for the subsystem with two identical spheres at different boundaries. The blue curve represent disjoint bulk hypersurfaces and the red curve represents the traversing configurations.}
\label{LRsphereS}
\end{figure}

We consider the holographic entanglement entropy for two identical spheres, $\p\S_{\mbox{\tiny L}}$ and $\p\S_{\mbox{\tiny R}}$ of radius $R$, anchored at different boundaries of the wormhole geometry. This setup admits the two types of configurations which are represented in Figure \ref{LRsphere}. The disjoint configuration, painted in blue on the left panel of Figure \ref{LRsphere}, is associated with two hypersurfaces in the bulk which probe the Left and Right side of the wormhole geometry, respectively. The traversing configuration, in red on the right panel of Figure \ref{LRsphere}, joins the spherical surfaces $\p\S_{\mbox{\tiny L}}$ and $\p\S_{\mbox{\tiny R}}$ with a hypersurface which traverses the wormhole, it is refracted in both bubbles (at Left and Right sides) and at the throat. 
To construct the traversing $\S$ configuration anchored at the two entangling surfaces of radius $R$, $\p\S_{\mbox{\tiny L}}$ and $\p\S_{\mbox{\tiny R}}$, one at each boundary, we require the matching condition of equation (\ref{mc}) at the throat and at both bubbles. As previously, the latter reads $\Pi|_{-} = \Pi|_{+}$ for a common orientation of the normal vectors at each side of the matching surface. 
Particularly, if we use expression (\ref{Pi_sphere}) to write $\Pi|_{\mbox{\tiny L,R}}$ at each side of the throat at $r_{\text{\tiny th}}=1$, we have
\be \label{mc_th}
\Pi_{\mbox{\tiny L,R}} 
= 
- \frac{
2 \r_{\mbox{\tiny th }}
+ 
\r'_{\mbox{\tiny th }} \, \lp 3 -2 \r_{\mbox{\tiny th }}^2 \rp
+ 
   3 {\r'_{\mbox{\tiny th }}}^3 \, (2 + \r_{\mbox{\tiny th }}^2 )
     }
     {3\ell^2\, \r_{\mbox{\tiny th }}^2 \, ({\r'_{\mbox{\tiny th }}}^2+1/3)^{3/2}} 
     \,,
\ee
with $\rho_{\mbox{\tiny th }} = \r_i(r_{\mbox{\tiny th }})$, $\rho'_{\mbox{\tiny th }} = \r'_i(r_{\mbox{\tiny th }})$, and $\Pi_{\mbox{\tiny L,R}} $ associated to the unit normal $n^\mu_{\mbox{\tiny L,R}}$ whose orientation is in the increasing direction of the $r$ coordinate of the Left or Right side, correspondingly. In terms of these quantities, the matching condition at the throat reads $\Pi|_{\mbox{\tiny L}} = -\Pi|_{\mbox{\tiny R}}$, with the sign corrected according to the common orientation. The identical surfaces $\p\S_{\mbox{\tiny L}}$ and $\p\S_{\mbox{\tiny R}}$ force to have a $Z_2$ symmetric traversing configuration which determines that $\Pi|_{\mbox{\tiny L}} =  0 = \Pi|_{\mbox{\tiny R}}$ at the throat. This condition fixes how the hypersurface $\S$ bends at the position of the throat for each possible value of its radius $\rho_{\mbox{\tiny th }}$, i.e. it fixes the value of $\rho'_{\mbox{\tiny th }}$ in terms of the size $\rho_{\mbox{\tiny th }}$ of $\S$ at the throat. There are three possible solutions for $\Pi|_{\mbox{\tiny L,R}} =  0$, each corresponding to a different boundary condition for $\S$ at the position of the throat. The latter are summarized in the graph of Figure \ref{solsLR}.
The three possible solutions are represented in Figure \ref{solsLR} with the red, brown and orange curves for $\rho'_{\mbox{\tiny th }}$ vs. $\rho_{\mbox{\tiny th }}$. The red curve represents solutions for $\S$ configurations which traverses the wormhole throat admitting any value of $\rho_{\mbox{\tiny th }}$, while the orange and brown solutions admit only $\rho_{\mbox{\tiny th }} \geq \r_{\mbox{\tiny th }}^c \simeq 3.9$. The red solutions correspond exclusively to negative values of $\rho'_{\mbox{\tiny th }}$, which imply a phyton lunch shape configuration, like the one shown on the right in Figure {\ref{LRsphere}}. Moreover, python lunch 
solutions attain the minimum value of the finite part of the HEE in comparison with those obtained using the orange and brown curve solutions of Figure \ref{solsLR}. For this reason, we will use phyton lunch configurations to compare the holographic entanglement entropy of the LR-traversing $\S$ with the non-traversing one corresponding to the disjoint spherical hypersurfaces (Left and Right blue spherical surfaces on the left in Figure \ref{LRsphere}). To do so we compute the finite part as
\be
S_{\text{\tiny finite}}
= \lim_{\e \to 0 
}
\lp
S_{\text{\tiny HEE}}^{\text{LR sphere}}
- 2 \, S_{\text{\tiny div}}^{\text{sphere}}
\rp \,
\ee
for each of the them, in terms of different values of the radius $R= \rho(r_{\infty})$. The results are plotted in Figure \ref{LRsphereS}. We found that for radii of the entangling surfaces larger than a critical length $R_c \simeq 0.45$ the LR traversing configurations have the minimal finite HEE. The dominant curve constructed by the minimal between both preserves the concavity of the physical curve of finite HEE vs $R$. The latter captures a phase transition in which a subsystem made up of the union of regions in different boundaries couple through the wormhole connecting them.

\subsubsection{HEE in the wormhole geometry: A slab volume of width $L$}

We now repeat the analysis for the volume of a slab in one boundary of the wormhole geometry. The slab has infinite dimensionless size $L_\perp^2$ in the $x_2$-$x_3$ directions and width $L$ in the $x_1\equiv x$ direction. The holographic hypersurface $\S$ is anchored at the surface of the slab and has an indued metric given, in coordinates $\lk r,x_2,x_3 \rk$, by the line element
\be
dh^2 = \ell^2 \lbr h_{rr}(r)\, dr^2 + r^2 (dx_2^2+ dx_3^2) \rbr 
\ee
where
\be
h_{rr}(r) = \frac{1}{f(r)} + \lp r\, {x'}(r) \rp^2  \ ,
\ee
with $x(r) \in [-L/2, L/2]$ and width given by
\be
\ell L = 2\ell \int_{r_0}^{r_\infty}  {x'}(r)  \, dr\,.
\ee
If  $\S$ is a U-shaped hypersurface shrinking in the outer or inner bulk geometry, then $r_0$ is the turning point. Refracted configurations appear for $r_0 < r_{\text{\tiny b}}$, for which we distinguish the solution $x_i(r)$ and $x_e(r)$ in the corresponding regions. If $x_i(r)$ reaches the throat, the stretched configurations for the slab case are constructed  analogously to what was done in the spherical case. For slab-like hypersurfaces $\S$ we define its space-like unit normal as
\be
s_{\m}  =
\frac{\ell \, r }{\sqrt{f(r) \, h_{rr}(r)}} 
\lp
- x'(r) \, \d_\m^r 
+ \d_\m^x
\rp \,.
\ee
The matching conditions 
at constant time and fixed $r$ are held over the hypersurface $\mathcal{H}$ of a bubble or at the throat.
The induced metric over $\p \S_{\mbox{\tiny R}}$, embedded in $\mathcal{H}$, has line element
\be \label{borde_pegado_slab}
d\g^2 = 
\ell^2 r^2 
\lp 
dx_2^2 +  dx_3^2
\rp \,,
\ee
where $r=r_{\text{\tiny th}}$ or $r=r_{\text{\tiny b}}$, while  the unit normal to $\p\S_{\mbox{\tiny R}}$ which is tangent to $\mathcal{H}$ is
\be
\eta_{\m} 
= \ell \, r \, \d_\m^x \ ,
\ee
and the unit normal which is tangent to $\S$ is given as
\be
n_{\m}  
=
\frac{\ell}{f(r) \,\sqrt{h_{rr}(r)}} 
\lp
{ \d^r}_\m 
+  r^2\, f(r)\, x'(r) \, {\d^x}_\m
\rp%
\,,
\ee
pointing in the increasing direction of the bulk coordinate $r$. 
The 2-dimensional extrinsic curvature of $\p \S_{\mbox{\tiny R}}$ with respect to $\eta^\mu$ is zero, and the extrinsic curvature with respect to $n_{\m}$ is
\be
\mathcal{K}^{(n)}_{\m \n}
=
 \frac{\ell \, r }{\sqrt{h_{rr}(r)}}
\,
\lp
\d^{x_2}_\m \, \d^{x_2}_\n 
+
\d^{x_3}_\m \, \d^{x_3}_\n 
\rp
\ee
which trace is given by
\be
\mathcal{K}^{(n)} 
=
  \frac{2}{\ell \, r \, \sqrt{h_{rr}(r)}} \,.
\ee
The 2-dimensional intrinsic Ricci scalar of $\p \S_{\mbox{\tiny R}}$ is null so, putting everything together, 
we have
\be \label{Pi_slab}
\Pi_{\mbox{\tiny slab}}
=
-
\frac{
\sqrt{f(r)} \,  x'(r) \,  \lbr r^2 +  ( x'^2(r)\, r^4 - 1 ) f(r) \rbr
}{
\ell^2\,r \, \lbr 1 + r^2 \, f(r) \, x'^2(r) \rbr^{3/2}
}
\ee
for slab-like configurations.

\begin{figure}[H]
\centering
\includegraphics[width=8cm]{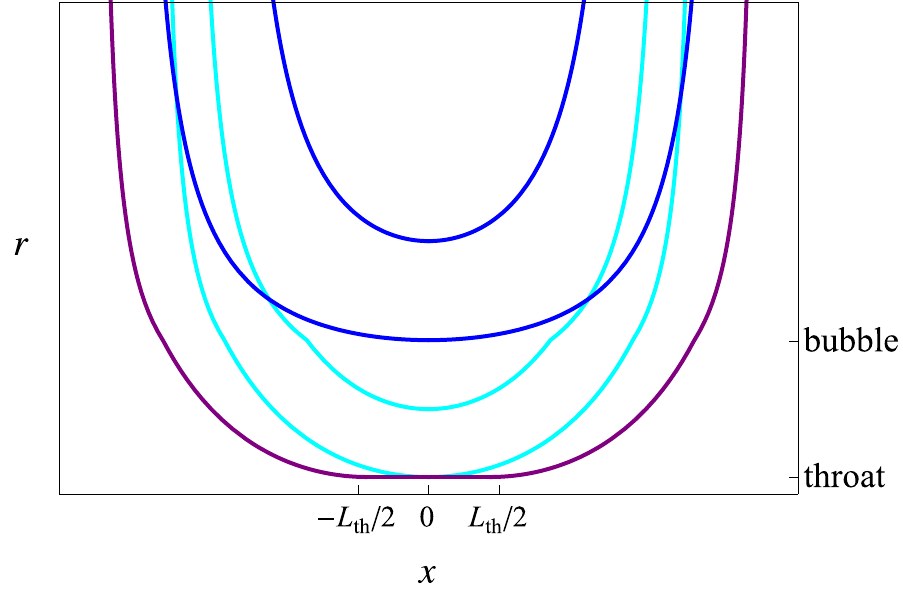}
\hspace{0.4cm}
\includegraphics[width=8cm]{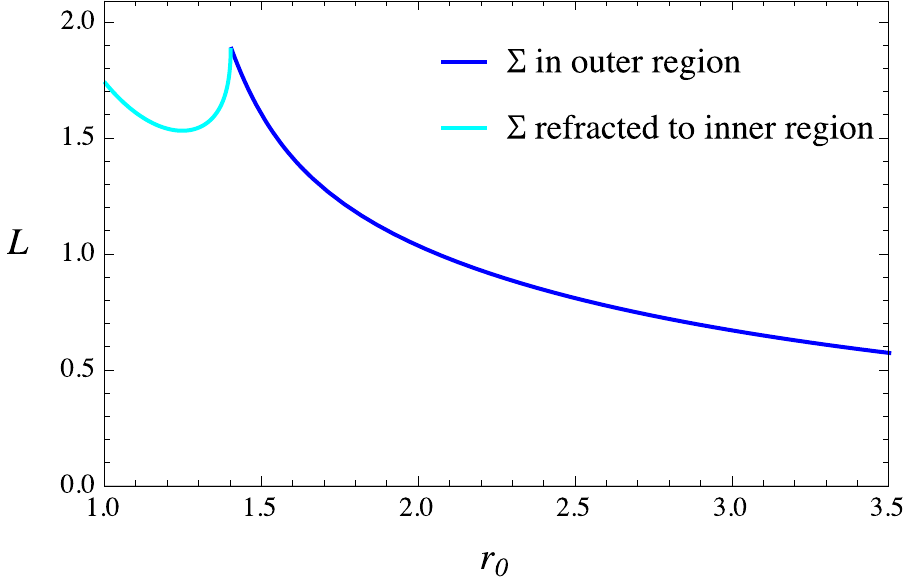}
\caption{Left: Shape of $\S$ hypersurfaces for slab-like configurations. Examples of $\S$ in the outer region (blue), refracted to the inner region (cyan) and for the case of a stretched $\S$ at the throat (purple) are shown in the left panel. Right: width $L$ vs. turning point $r_0$. 
}
\label{}
\end{figure}

\begin{figure}[H]
\centering
\includegraphics[width=8.3cm]{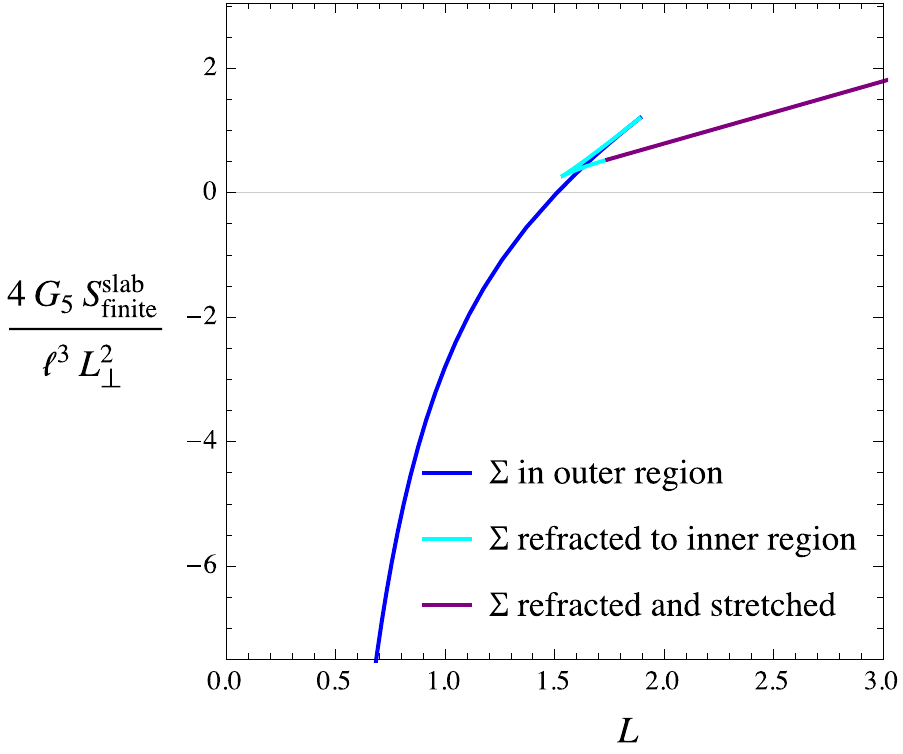}
\hspace{0.4cm}
\includegraphics[width=8.35cm]{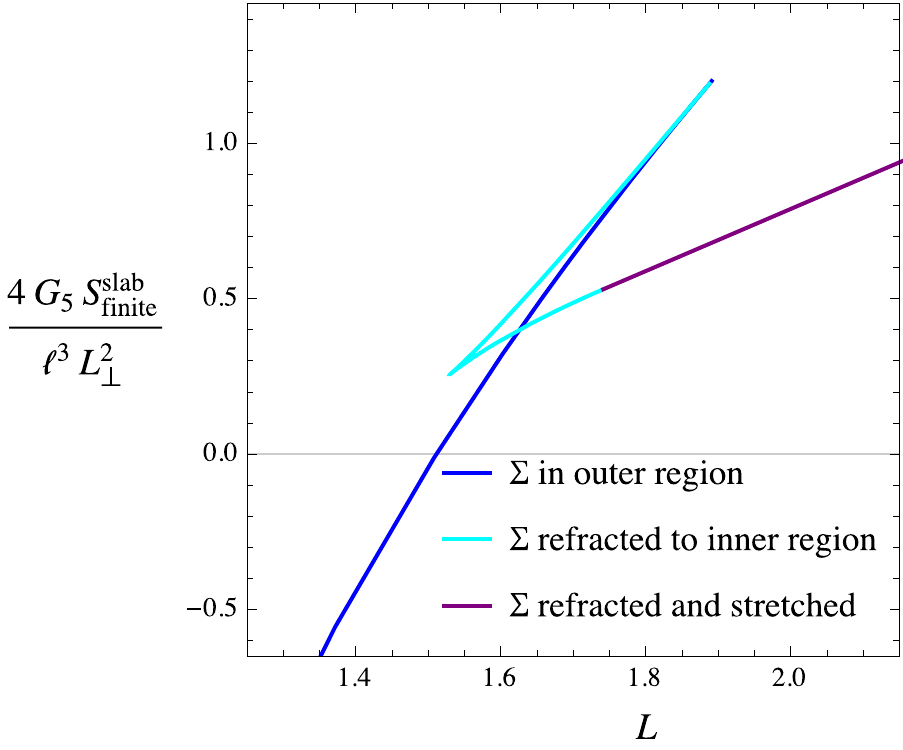}
\vspace{-0.25cm}\caption{$S_{\mbox{\tiny finite}}^{\mbox{\tiny slab}}$ vs. width $L$ for the subsystem given by a slab of size $\ell L$ on one boundary.} 
\label{S_slab_wh}
\end{figure}

The HEE functional for the slab in the wormhole geometry is
\be
S_{\text{\tiny HEE}}^{\text{slab}} 
= \frac{\ell^3}{4G_5}\, 
2 \, L_\perp^2
\int_{r_0}^{r_{\infty}}
dr \,
{\mathcal{L}(x'(r),r)} \,,
\ee
with the 
one-dimensional Lagrangian density
\be
{\mathcal{L}(x'(r),r)}
=
\frac{
r^2 + f(r) + r^4 \, f(r)\, x'^2(r)
}{
\sqrt{ f(r) \lbr 1+  r^2 \, f(r) \, x'^2(r) \rbr}
} \ ,
\ee
evaluated in the corresponding -inner or outer- region. In the case of a stretched configuration we have 
\bea
\lb{S_slab_str} 
S_{\text{\tiny HEE}}^{\text{slab (stretched)}} 
&=& \frac{\ell^3 }{4G_5}\, 2 \, L_\perp^2
\lp
\int_{r_{\mbox{\tiny th}}}^{r_\infty}
dr  \,
\mathcal{L}(x'(r),r)
+
r_{\mbox{\tiny th}}^3 \int_{0}^{L_{\mbox{\tiny th}}/2}
\, dx
\rp
 \nn \\
&=& \frac{\ell^3 }{4G_5}\, 2 \, L_\perp^2
\int_{1}^{r_\infty}
dr  \,
\mathcal{L}(x'(r),r)
+
 \frac{\ell^3 \, L_\perp^2 L_{\mbox{\tiny th}}}{4G_5}\,  
\eea
with $L_{\mbox{\tiny th}} = 2|x_i(r_{\mbox{\tiny th}})|$ the stretched segment at the throat. 
In every case, for asymptotic values of $r$, the integrand behaves as 
$\mathcal{L}(\rho,\rho'(r),r) \sim
2 \, r$, producing the expected UV-divergent terms 
$S_{\text{\tiny div}}^{\text{slab}}$,
as in (\ref{S_div_slab}).
The finite terms, independent of the UV-cutoff, are obtained as the regularized quantity 
\be \lb{finite_slab_wh}
S_{\text{\tiny finite}}^{\text{slab}}
= 
\lim_{\e \to 0} 
\lp
S_{\text{\tiny HEE}}^{\text{slab}}- S_{\text{\tiny div}}^{\text{slab}}
\rp
\ee
with $\e = 1/r_\infty$. The results of the computation are plotted in Figure \ref{S_slab_wh} as the finite HEE against $L$.
For $L \ll 1$ the leading finite, cutoff independent, behavior is
$S_{\mbox{\tiny finite}}^{\mbox{\tiny slab}} \sim - 
L^{-2}$, 
as found in pure AdS. The slope of the linear part of the blue curve, where the finite HEE turns positive, is approximately ${4G_5}S_{\mbox{\tiny finite}}^{\mbox{\tiny slab}}/(\ell^3L_{\perp}^2 L) \simeq  2.6$, while the slope of the linear curve in purple is $1$.  For $L \simeq L_{\mbox{\tiny th}}\gg 1$ 
the finite HEE scales as the volume of the stretched slab at the throat, which in this limit is approximately the volume of the region at the boundary.
The right panel in Figure \ref{S_slab_wh} shows a cusp in the transition from outer to inner solutions of the configuration space, while a smooth transition is found if the size increases to change from inner to stretched surfaces in the bulk.

\subsubsection{HEE in the wormhole geometry: Two slab volumes of width $L$}

\begin{figure}[H]
\centering
\includegraphics[width=8.6cm]{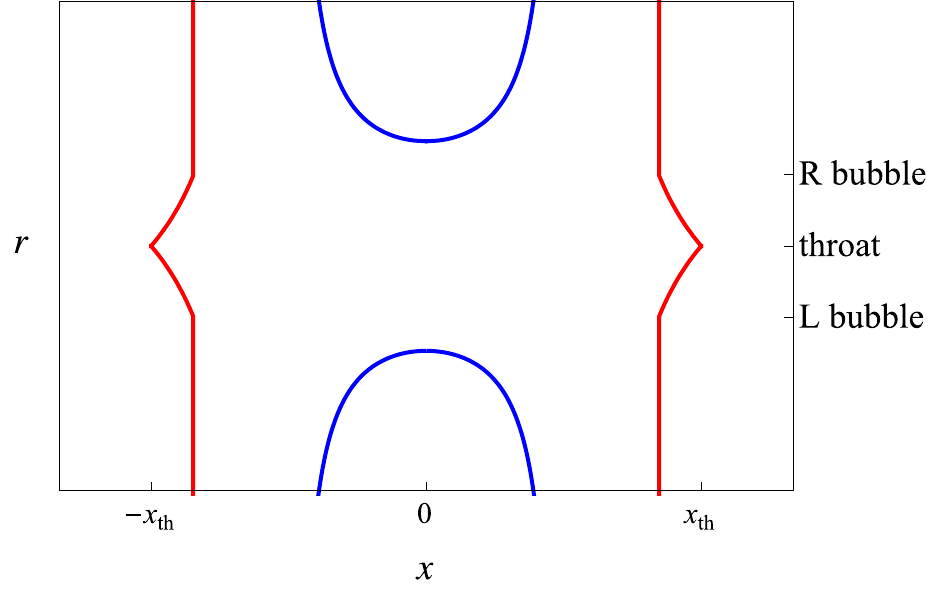}
\caption{In blue: Left and Right side disjoint slab hypersurfaces in the outer regiones of the bulk wormhole geometry. In red: Left-Right slab-like hypersurface traversing the the wormhole throat.}
\label{LRslabs}
\end{figure}

We consider the HEE for the system of two identical slab-like volumes anchored at different boundaries. This setup allows two types of surface configurations in the bulk, as shown in Figure \ref{LRslabs}.
The blue disjoint configuration is associated with two copies of the slab surface of width $L$, one in each boundary, probing the Left and Right side of the bulk geometry, respectively. The red 
configuration connects the Left and Right boundaries with a hypersurface traversing the wormhole, which is refracted in both bubbles (Left and Right) and at the throat.

To construct the traversing configuration we requiere the matching condition $ [\Pi]=0$ at the throat. Particularly, to have a $Z_2$ symmetric configuration we need $\Pi_{\mbox{\tiny L,R}} = 0$ at each side of the matching surface of the throat, i.e. a vanishing value of (\ref{Pi_slab}) at $r= r_{\text{\tiny th}}=1$. This condition fixes how the hypersurface $\S$ bends at the position of the throat. There are three possible values for the slope $x'(r_{\text{\tiny th}})$ which satisfy the vanishing of (\ref{Pi_slab}), these are;
\be
v_0=0\, ,
\qquad
v_1=\sqrt{2/3}\, \quad \mbox{and} \quad
v_2=-\sqrt{2/3} \,.
\ee
These solutions are independent of the width of the slab-like hypersurface $\S$, and the LR configurations which minimize the HEE are obtained using the values $v_1$ or a $v_2$ for $x'(r_{\text{\tiny th}})$, which produce an inner solution with derivative given by
$$
x_{i}'(r)= \pm \frac{\sqrt{f_i(r)-r^2}}{r^2\, \sqrt{f_i(r)}}
$$
for the interior part $\S^{\text{\tiny i}}$ of $\S$\footnote{If $v_0=0$ is used, the inner solution has $x_{i}'(r)=0$, identically. This means $\Pi_{\mbox{\tiny slab}}|_{r=r_{\mbox{\tiny b}}^{\mp}}=0$ and then $x_{e}'(r)=0$, i.e., the LR traversing slab is a straight hypersurface from one boundary to the other. Nevertheless, the latter has larger value of the finite HEE than the one selected here.}. 
Using $x_{i}'(r)$ we have that $\Pi_{{\mbox{\tiny slab}}}|_{r=r_{\mbox{\tiny b}}^-}=0$ at the matching with the outer part $\S^{\text{\tiny e}}$ and, consequently, we must have $\Pi_{{\mbox{\tiny slab}}}|_{r=r_{\mbox{\tiny b}}^+}=0$. The latter produces $x_{e}'(r)=0$ for $r>r_{\mbox{\tiny b}}$ as the only compatible solution. The computation of the finite part of the HEE with this type of LR slab-like hypersurface gives
\be \label{S_finite_LRslab}
S_{\text{\tiny finite}}
= \lim_{\e \to 0 
}
\lp
S_{\text{\tiny HEE}}^{\text{LR slab}}
- 2 \, S_{\text{\tiny div}}^{\text{slab}}
\rp 
\simeq
-\frac{\ell^3 \, L_{\perp}^2}{4G_5} \, 0.64
\ee
which is smaller than the one obtained with disjoint (blue) surfaces for $L$ grater than the critical length $L_c \simeq 1.4$ and independent of the length of the slabs. The latter is analogous to what happens in phase transitions of confining backgrounds in which there is a point where bulk surfaces end. Here, for $L > L_c$, the physically favorable finite HEE associated to the disconnected subsystem made up of the union of two slab-like regions in different boundaries is given by the LR traversing configurations, which captures a deep infrared behaviour given by the non-vanishing uniform term (\ref{S_finite_LRslab}), characteristic of the wormhole background.

The results obtained in this paper can be easily generalized to other configurations such as disconnected 3-volumes in the same copy of the boundary theory by applying the same techniques we have employed, for example, as was done in \cite{BA2014} using multiple slabs. This yields a richer phase diagram with phase transitions in which the separation scales competes with the presence of the wormhole. 


\section*{Acknowledgements}

The authors thank David Blanco, Alberto G\"uijosa, Guillem P\'erez-Nadal for discussions. M.C. thanks Tom\'as Andrade for his comments on the manuscript. The work of M.C. is partially supported by Mexico's National Council of Science and Technology (CONACyT) grant A1-S-22886, DGAPA-UNAM grant IN107520 and PASPA fellowship. The work of G.G. and E.R.d.C. was supported by CONICET and ANPCyT through grants PIP1109-2017, PICT-2019-00303.


\end{document}